\documentclass[10pt, twocolumn,aps,floatfix,showpacs,prb]{revtex4-1}

\usepackage{amsmath}
\usepackage{amssymb}
\usepackage{amsfonts, epsfig}
\usepackage{graphicx}
\usepackage{setspace}
\usepackage{calc}
\usepackage{floatflt}

\def\H{{\cal H}}
\def\e{\varepsilon}

\def\ket#1{|#1\rangle }
\def\bra#1{\langle#1 | }

\def\expect#1{\langle#1 \rangle}
\def\correl#1#2{\ll\!\!#1|#2\!\!\gg(z)}
\def\non{\nonumber \\ }
\def\w{\omega}
\def\punkt{\;\; .}
\def\Tr#1{\textrm{Tr}\left[#1\right]}

\def\lph{\lambda_{ph}}
\begin{document}

\title{Influence of vibrational modes on the quantum transport
through a nano-device}

\author{Andre Jovchev}
\author{Frithjof B.~Anders}
\affiliation{Lehrstuhl f\"ur Theoretische Physik II, Technische Universit\"at Dortmund, 44221 Dortmund,Germany}

\date{\today}

\begin{abstract}
We use the recently proposed scattering states numerical
renormalization group (SNRG) approach to calculate  $I(V)$ and the
differential conductance through a single molecular level coupled to a
local molecular phonon.  We also discuss the equilibrium physics of
the model and demonstrate that the low-energy Hamiltonian is given by
an effective interacting resonant level model.  From the NRG level
flow, we directly extract the effective charge transfer scale
$\Gamma_{\rm eff}$ and the dynamically induced capacitive coupling
$U_{\rm eff}$ between the molecular level and the lead electrons which
turns out to be proportional to the polaronic energy shift $E_p$  for
the regimes investigated here. The equilibrium spectral functions for
the different parameter regimes are discussed. The additional phonon
peaks at multiples of the phonon frequency $\w_0$ correspond to
additional maxima in the differential conductance. Non-equilibrium
effects, however, lead to significant deviations between a symmetric
junction and a junction  in the tunnel regime.  The suppression of the
current for asymmetric junctions with increasing electron-phonon
coupling, the hallmark of the  Franck-Condon blockade, is discussed
with a simple framework of a combination  of (i) polaronic level
shifts and (ii) the effective charge transfer scale $\Gamma_{\rm
  eff}$. 
\end{abstract}


\maketitle

\section{Introduction}

In the quest for size-reduced and possible low-power consuming
electronic devices the proposal\cite{AviramRatner1974}  of using
molecular junctions for electronics has sparked a  large interest in
understanding the influence of molecular vibrational modes onto the
electron charge transfer through a molecule.  In the simplest building
block of molecular electronics,  a molecule  is connected to two
leads. The non-linear current through such a device can be controlled
by an external gate tuning the molecular
levels.\cite{Chen1999,Donhauser2001}  In some cases a sudden drop of
the current has been observed  with increasing bias
voltage\cite{Chen1999} which translates into a negative differential
conductance. Also, hysteretic behavior of the I(V)
curve\cite{LiHysteresis2003} has been  reported when sweeping the
voltage with a very  small but finite rate. It has been suggested that
such a reduction of conductance and the hysteretic behaviour might
originate in conformational changes in these complex
molecules.\cite{Donhauser2001} Vibrational coupling has also been
found of importance in break junctions\cite{TalBreakJunction2008} and
suspended carbon nanotube quantum
dots. \cite{SapmazCarbonNanotubes2005,SapmanPRL2006,PopEtAl2005,NatureFranckCondonCarbonNanotubes2009}
An excellent review\cite{GalperinRatnerNitzan2007}  by Galperin et
al.\ summarizes comprehensively the different theoretical approaches and experimental findings.

The theoretical description of such molecular junctions focuses only
on  those  molecular levels and vibrationals modes which are relevant
for the transport. In its simplest  version\cite{GalperinRatnerNitzanHysteresis2004,GalperinRatnerNitzan2004,GalperinRatnerNitzan2007}
a single level coupled to a local Holstein phonon has been
considered. Typically rate equations\cite{KochOppen2005} or
Keldysh-Green function
approaches\cite{GalperinRatnerNitzan2007,KochFehske2011}  have been
applied to this problem.  One either expand the self-energy in powers
of the  electron-phonon coupling in the weak-electron phonon coupling
regime\cite{GalperinRatnerNitzan2007} or start from the exact solution
of the local problem using the Lang-Firsov
transformation\cite{LangFirsov1962} and expand in powers of the
tunneling matrix element.\cite{KochFehske2011}  The polaron formation
on a molecular wire  as a mechanism for negative differential
resistance\cite{SapmazCarbonNanotubes2005,SapmanPRL2006,PopEtAl2005}
which has been proposed  uses a simple mean-field
approximation.\cite{GalperinRatnerNitzanHysteresis2004} Using the
imaginary-time formalism,\cite{HanHeary2007,HanDirksPruschke2012}
however, the I-V curve  of single orbital molecular junction does not
show phononic site peaks.\cite{HanMolecule2010} Whether  this result
prevails when the fit-function\cite{HanMolecule2010} for the
electronic self-energy is replaced by the exact solution remains an
open question. Recently,  the iterative path-integral approach has
also been successfully applied\cite{HutzenEgger2012} to calculate
quantum transport  for moderate and high temperatures compared to the
charge-transfer rate $\Gamma_0$. 

In this paper, we will briefly review the known  physics of such a polaronic model from a
renormalization group perspective.  
The  low-energy Hamiltonian of the  minimal model for molecular devices is given by 
a effective interacting-resonant level model:\cite{Schlottmann1980,MethaAndrei2005,EidelsteinSchiller2012}
a considerably large Coulomb repulsion $U_{\rm eff}$ is dynamically generated  which governs the
zero-bias transport as function of the gate voltage\cite{EidelsteinSchiller2012}  as well as the shape
of the spectral functions, as we will demonstrate in our paper.  We
will present a detailed scaling analysis of the  renormalized
charge-transfer rate  $\Gamma_{\rm eff}$ and $U_{\rm eff}$ in  the
weak to intermediate coupling regime  which covers a complementary
regime of the once  studied recently by Eidelstein et
al.\cite{EidelsteinSchiller2012}  We extend the discussion of the
equilibrium properties to the single-particle spectral functions which
governs the transport in the tunneling regime.  Using the scattering
states numerical renormalization group
(SNRG)\cite{AndersSSnrg2008,SchmittAnders2010,SchmittAnders2011} we
present the non-perturbative results for  I-V characteristics of the
model far from equilibrium at low temperature augmenting recent
studies  of other non-perturbative approaches to larger temperatures.\cite{HutzenEgger2012} 

Focusing on a  single vibrational
mode,\cite{GalperinRatnerNitzan2007,KochOppen2005,HaertleThoss2011,HutzenEgger2012,EidelsteinSchiller2012}
the equilibrium physics of  two extreme limits have been well
understood. 

In the adiabatic limit, where the phonon frequency is the smallest energy scale
of the problem a small electron-phonon coupling yields a reduction
of the phonon frequency by particle-hole excitations. In leading order, the  correction
to  the electronic self-energy is quadratic in the coupling constant.   
This limit has been pioneered  by Caroli et al.\cite{Caroli71,*Caroli72}
in the context of tunnel junctions and applied to molecular junctions using
the self-consistent Born approximation.\cite{GalperinRatnerNitzan2004} 

In the opposite limit for very small tunneling rates $t_\alpha$, one starts from
the exact solution of the local  problem, $t_\alpha=0$, by applying a Lang-Firsov 
transformation.\cite{LangFirsov1962,Mahan81} A displaced phonon with
an unrenormalized phonon frequency $\w_0$ and a polaron with a
shifted  single-particle energy  is formed locally. When tunneling,
the electron has to be extracted from the polaronic quasi-particle
which can be done at many difference excitation energies differing by
multiples of the phonon frequency. Each pole  of the single-particle
Green function contributes only with a fractional weight to the
spectra.\cite{Mahan81} If the phonon energy $\w_0$ is large compared
to the electron energy scales, the bosonic mode can be considered in
its ground state, which leads to an exponential renormalization of the
tunneling matrix element  $t_\alpha\to t_\alpha\exp(-g^2/2)$, where
$g=\lph/\w_0$  and $\lph$ denotes the electron-phonon interaction
strength. In this anti-adiabatic limit, the strong electron-phonon
coupling yields a polaronic shift  of the single particle level and,
depending on the bare parameters, a reduction of the tunneling rate.
This leads to the Franck-Condon blockade in quantum
transport.\cite{KochOppen2005,GalperinRatnerNitzan2007,NatureFranckCondonCarbonNanotubes2009,HutzenEgger2012}

In the limit of vanishing charge transfer rates, i.\ e.\
$(\Gamma_0/\w_0), (\Gamma_0/\lph)\ll 1$, the I-V characteristic
exhibits rather sharp steplike features  which are a reminiscence  of
the exact solution of the local spectral
function\cite{LangFirsov1962,Mahan81}  as  recently obtained using
kinetic \cite{LeijnseWegewijs2008} equation, equation of motion decoupling schemes\cite{Flensberg2003}  or master
equations\cite{HaertleThoss2011} or a Keldysh Green
function\cite{Haertle2011} approach.  Within such an approach
negative differential resistance is found in regimes which are
complementary to our scattering states NRG approach  used in this
paper. Such steplike features  with negative differential resistance
have been observed in suspended carbon nanotube
experiments.\cite{SapmazCarbonNanotubes2005,SapmanPRL2006}

The problem becomes more difficult  in the crossover regime between
the adiabatic and the non-adiabatic regime. Often simple rate
equations or self-consistent Born
approximation\cite{GalperinRatnerNitzan2007}  fail to describe the
proper renormalization of the parameters. Recently, Wilson's numerical
renormalization group (NRG)
approach\cite{Wilson75,BullaCostiPruschke2008}  has been adapted to
this problem.\cite{HewsonMeyer02}   A thorough
study\cite{EidelsteinSchiller2012}  has demonstrated the power of
this non-perturbative approach to reveal the interplay between the
different energy scales of the problem  in the crossover  regime. An
extended anti-adiabatic regime has been identified where the bare
charge transfer rate $\Gamma_0$ exceeds the phonon frequency but
remains below the polaron shift $E_p=g^2\w_0$. The scaling behaviour
of renormalized charge  transfer rate $\Gamma_{\rm eff}$ was obtained
as function of the $g$ and the polaron shift $E_p$.

Galperin et al.\cite{GalperinRatnerNitzanHysteresis2004}
discuss the possibility of polaron formation on a molecular wire  as a
mechanism for negative differential resistance (NDR) uses a simple
mean-field approximation.  Such hysteresis arise for branch cuts of
the the non-linear equations at sufficiently larger coupling.   It
remains unclear whether this effect survives in the exact solution of
the problem.   A prominent example of a false hysteresis in the I-V
characteristics is the conserving GW-approximation to Anderson model
out of equilibrium as shown in a paper by Spataru et al.\
\cite{spataruGwMillis2009} 
In a recent paper it has been proposed\cite{MuehlbacherThoss2012} 
that bistability  signatures in non-equilibrium charge transport through 
molecular quantum dots might be linked to subtle differences in the initial conditions.

\subsection{Plan of the Paper}

The paper is organized as follows. We start with a definition of the model
considered in this paper and relate our form of the electron-phonon
coupling to choices in the literature in Sec.\ \ref{sec:anderson-holstein-model}.
Decoupling the local level from the two leads, the local dynamics can be solved
exactly by the well-known Lang-Firsov transformation\cite{LangFirsov1962} 
summarized in Sec.\ \ref{sec:lang-firov}. We briefly review
the numerical renormalization group  (NRG) approach to quantum impurities
and the scattering states NRG used for the quantum-transport calculations in
the Secs.\ \ref{sec:nrg}  and \ref{sec:snrg}. 

The Sec.\ \ref{sec:equi-nrg} is devoted
to present the equilibrium NRG results. We extract the parameters of the effective
low-energy Hamiltonian in the regime $\w_0>\Gamma_{\rm eff}$ which is given by
an interacting resonant level model (IRLM). After summarizing the technical details
how to extract the effective parameter directly from the NRG level flow, we discuss
the scaling properties of $\Gamma_{\rm eff}$  and $U_{\rm eff}$ as function of the bar
model parameters in Sec.\ \ref{sec:NRG-renormalized-params}.

The equilibrium spectral function already reveals important
information for the quantum-transport since it is proportional to the
transfer-matrix in the tunneling regime. Therefore, we present
equilibrium spectral functions for symmetric and asymmetric junctions
as well as benchmarks for our non-equilibrium Green function algorithm
in Sec.\ \ref{sec:equilibrium-rho}. 

Sec.\ \ref{sec:quantum-transport} is devoted to the results for the
non-equilibrium quantum transport. We present our data for symmetric
and asymmetric junctions for different phonon frequencies and junction
asymmetries. We comment on the observation  of the Franck-Condon
blockade physics and the recovery of the tunneling limit. We conclude
with a short discussion and an outlook  in Sec.\
\ref{sec:discussion-outlook}.

\section{The Anderson-Holstein model}

\subsection{Definition of the model}
\label{sec:anderson-holstein-model}

\begin{figure}[htbp]
\begin{center}
\includegraphics[width=75mm]{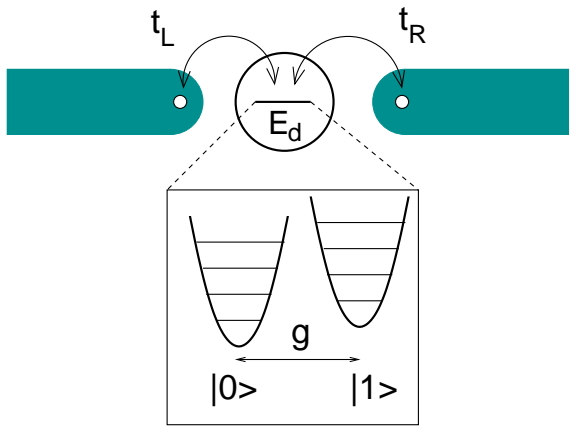}

\caption{(Color online.) Minimal model of a single-molecular level with energy $E_d$ 
coupled to two leads with tunneling matrix elements $t_L$ and $t_R$.
Depending on the local charge configuration $\ket{0}$ or $\ket{1}$,
the Holstein phonon ground state is shifted. The relative displacement between
the two configuration is given by the dimensionless electron phonon coupling $g=\lph/\w_0$.
The phononic excitations for a fixed charge are multiples of the oscillator energy $\w_0$.
}
\label{fig:1-model}
\end{center}
\end{figure}

The minimal, non-trivial model for molecular-electronics
\cite{GalperinRatnerNitzan2007,HutzenEgger2012}
comprises a single spinless level whose charge is coupled locally to
a single Holstein phonon stemming from the dominating
vibrational mode of the molecule. For the 
current transport this level is coupled to the two leads which
are often considered as featureless bands for simplicity. 
In real materials, band features are important but only influence the 
single-particle properties which can be accounted for in a frequency dependent
charge transfer rate $\Gamma(\w)$.  

This spinless Anderson-Holstein model is depicted schematically in
Fig.\ \ref{fig:1-model}. Depending on the local charge configuration,
the local harmonic oscillator is displaced and the distance between
the two ground states is given by $g$ (see below.) At the
particle-hole symmetric  point and in absence of a coupling to the
leads,  the displaced oscillator  ground states are given by the two
coherent states $\ket{\pm g/2}$,  and we immediately can understand
the underlying Franck-Condon physics by the suppression of the overlap
$\bra{-g/2}g/2\rangle = \exp(-g^2/2)$ of the two ground states with
increasing electron-phonon coupling $g$. 

The spin-less two lead resonant level model (RLM) defined by the Hamiltonian
\begin{subequations}
\label{equ:one}
\label{eqn:two-lead-RLM}
\begin{eqnarray}
H_0 &=& H^0_{imp} + H_{T}+ H_{leads}\\
H^0_{imp} &=& E_d d^\dagger d
\\
H_{T} &=& 
\sum_{\alpha=L,R} \frac{t_\alpha}{\sqrt{N}} \sum_{k} (d^\dagger c_{k\alpha} + c^\dagger_{k\alpha} d)
\\
 H_{leads}&=& \sum_{\alpha=L,R} \sum_{k} \e_{k\alpha}c^\dagger_{k\alpha}c_{k\alpha}
\end{eqnarray}
\end{subequations}
is often used\cite{GalperinRatnerNitzan2007,HutzenEgger2012} as the
simplest model to describe  quantum transport through  a molecule or a
suspended
nano-bridge.\cite{MolecularElectronicsBook2005,NatureFranckCondonCarbonNanotubes2009}
$d(d^\dagger)$ annihilates(creates) an electron on the device with
energy $E_d$, and $c^\dagger_{k\alpha}$ creates an electron in the
lead $\alpha$  with energy $\e_{k\alpha}$.  The local charge-transfer
rate   to each lead $\alpha$  is given by $\Gamma_\alpha=\pi
t_\alpha^2 \rho_\alpha(0)$, where $\rho_\alpha(\w)$ is  the  density
of states of lead $\alpha$. Throughout the paper we will use a lead
independent constant density of states  $\rho_\alpha(\w) = (1/2D)
\Theta(D-|\w|)$   for simplicity;  $D$ denotes the band width of both leads.

We only account for a single local vibrational mode with energy $\w_0$ created by 
$b^\dagger$ 
whose dimensionless displacement operator
$\hat x=(b^\dagger + b)$ is coupled to the density $\hat n_d=  d^\dagger d $
of the  local level via 
\begin{eqnarray}
\label{eq:H-ph}
H_{ph} &=& \w_0 b^\dagger b + \lambda_{ph}  (b^\dagger + b)\left( \hat n_d - \frac{1}{2}\right)
\,\, .
\end{eqnarray}
The form of the interaction in (\ref{eq:H-ph})
ensures particle-hole  (PH) symmetry for $E_d=0$ and particle-hole symmetric leads.

Focusing only on the local dynamics defined by 
$H_{imp} = H^0_{imp} + H_{ph}$,
the harmonic oscillator will be shifted upon changing of the local occupancy away 
from half filling. This can be made explicit by introducing the arbitrary constant $n_0$
and making the trivial substitution
\begin{eqnarray}
d^\dagger d - \frac{1}{2}&=&  (d^\dagger d -n_0) +(n_0- \frac{1}{2})
\punkt
\end{eqnarray}
Then
$H_{imp}$ takes the exact form
\begin{eqnarray}
\label{eqn:shifted-oscillators}
H_{imp}&=&  \w_0 \bar b^\dagger \bar b + \lambda_{ph}  
(\bar b^\dagger + \bar b)\left( d^\dagger d - n_0\right)
\non
&&
+\bar E_d d^\dagger d  + 2E_p (n_0 -\frac{1}{2})^2
\end{eqnarray}
where the displaced phonon is created by $\bar b^\dagger$ defined as 
\begin{eqnarray}
\label{eqn:shifted-boson}
\bar b^\dagger &=& b^\dagger + g(n_0 -\frac{1}{2})
\, ,
\end{eqnarray}
and $g=\lph/\w_0$. The new single particle energy $\bar E_d$
\begin{eqnarray}
\label{eq:shifted-Ed}
\bar E_d &=& E_d + 2\frac{\lambda^2_{ph}}{\w_0}(\frac{1}{2}-n_0)
\end{eqnarray}
contains the polaronic energy shift $E_{p}=
\frac{\lambda^2_{ph}}{\w_0}=\w_0g^2$.  This polaron shift $E_p$ plays
an important role of defining the different
regimes\cite{EidelsteinSchiller2012}  of the  model in addition to the
dimensionless coupling constant  $g$ and the charge transfer rates $\Gamma_\alpha$.

In a mean-field decoupling of the electron-phonon interaction, $n_0$
would be replaced by self-consistent local charge expectation
$n_d=\expect{d^\dagger d}$ in Eqs.\
(\ref{eqn:shifted-oscillators}-\ref{eq:shifted-Ed}). A positive
(negative) $E_d$ leads to a depletion (filling up)  of the local level
where the  effective level $\bar E_d$ is further shifted to higher
(lower) energies    by a term proportional to  $E_{p}$. 

This has a profound impact on the zero-bias conductance as function of the detuning of the 
level $E_d$ by an external gate voltage. The local average occupation $n_d$ 
and therefore, the total displace charge $\Delta N$ due the presence
of the  impurity\cite{Langreth1966,Anders1991}  is initially
controlled by the ratio $E_d/\Gamma_{\rm eff}$, where $\Gamma_{\rm
  eff}$ is the low temperature charge fluctuation scale in the
Fermi-liquid fixed point. In the strong coupling regime, $g\gg 1$, the
effective level $\bar E_d $ becomes almost independent of the gate
voltage controlling the barelevel $E_d$. Hence, the occupation will be
nearly independent of $E_d$ and, therefore,  the conductance will have
a plateau unless $|E_d|$ exceeds the polaron shift $E_{p}$.  The
zero-bias conductance depicted in Fig.\ 13 of Ref.\
\onlinecite{EidelsteinSchiller2012} is a direct consequence of Eq.\ (\ref{eq:shifted-Ed}).

We also can use the parametrization of $H_{imp}$  with $n_0$ as
defined in Eq.~(\ref{eqn:shifted-oscillators}) to connect  $H_{imp}$
with  the  local  molecular Hamiltonian $H_m$
\begin{eqnarray}
\label{eqn:Hm-egger}
H_{m}&=&  \w_0  b^\dagger  b + \lambda_{ph}  
( b^\dagger + b)  d^\dagger d  +\e d^\dagger d
\end{eqnarray}
often used in the literature. \cite{ GalperinRatnerNitzanHysteresis2004, GalperinRatnerNitzan2007, HutzenEgger2012}
Setting $n_0=0$, $\e = E_d +E_{p}$ and neglecting the ground state energy shift,
$H_{imp}$ becomes identical to  $H_m$.

Keeping $\e$ fixed  in $H_m$ and increasing the electron-phonon coupling $\lph$ as in
Ref.\ \onlinecite{HutzenEgger2012} translates into a renormalization 
of the single-particle energy by a polaron shift $E_d= \e-E_{p}$ in $H_{imp}$.
Starting from $\e=0$, this implies a population increase of the fermion level upon increase
of $\lph$. In this parametrization it becomes apparent that $\lph$ induces a detuning away
from particle-hole symmetry.

Combining the RLM as given in Eq.\ (\ref{equ:one}) with a coupling to the local Holstein phonon
to $H=H_0 + H_{ph}$ defines the spinless Anderson-Holstein model.

\subsection{Lang-Firsov transformation}
\label{sec:lang-firov}

This Anderson-Holstein model is well studied 
\cite{HewsonMeyer02,GalperinRatnerNitzan2007,HutzenEgger2012,EidelsteinSchiller2012} 
and a text-book example for an exactly solvable
model\cite{Mahan81} in the limit $t_\alpha=0$. 
Already in the 1960s, it was shown  that the local Hamiltonian $H_{imp}=H^0_{imp}+H_{ph}$
can be  exactly diagonalized
using a Lang-Firsov 
transformation,\cite{LangFirsov1962}
\begin{eqnarray}
\hat U_{LF}&=& e^{-g(b^\dagger - b)(d^\dagger d -\frac{1}{2})}
\punkt
\end{eqnarray}
The new elementary excitations are
polarons annihilated by the operator $\bar d$
\begin{eqnarray}
\bar d &=& \hat U_{LF}^\dagger \, d \, \hat U_{LF} = e^{-g(b^\dagger - b)} d
\end{eqnarray}
and a free phonon with  unrenormalized  energy $\w_0$.  The polaron
shift $E_{p}$   enters  the ground state energy.   Acting on the
bosonic vacuum, the operator  $\hat U_{LF}=e^{-g(b^\dagger -
  b)(d^\dagger d -\frac{1}{2})}$ generates a coherent state  $\ket{\pm
  g/2}$ which describes the ground state of a  displaced harmonic
oscillator by the dimensionless displacement $\Delta x= \pm g/2$
depending on whether a local charge was present or absent (see also
the schematic figure \ref{fig:1-model}.)  This reflects the underlying
Franck-Condon physics extensively discussed in the
literature.\cite{KochOppen2005,GalperinRatnerNitzan2007,NatureFranckCondonCarbonNanotubes2009}

Neglecting the ground state energy shift, the total Hamiltonian $H'$ of the impurity
coupled to the leads  is given by
\begin{eqnarray}
\label{eqn:h-prime}
H' &=& H'_{imp} + \sum_{\alpha=L,R} \sum_{k} \e_{k\alpha}c^\dagger_{k\alpha}c_{k\alpha}
\non
&&
+
\sum_\alpha \frac{t_\alpha}{\sqrt{N}} \sum_{k} (e^{ g(b^\dagger - b)} \bar d^\dagger c_{k\alpha} + c^\dagger_{k\alpha} \bar de^{-g(b^\dagger - b)})
\non
H'_{imp} &=& E_d  \bar d^\dagger \bar d + \w_0 b^\dagger b 
\end{eqnarray}
after the unitary transformation $H'=\hat U^\dagger_{LF}H\hat U_{LF}$.
We have dropped the bar 
from the transformed phonon $\bar b = \hat U^\dagger_{LF} b \, \hat U_{LF}$ in 
Eq.\ (\ref{eqn:h-prime})  to keep
the notation simple. While the local Hamiltonian $H'_{imp}$ became simple and diagonal, 
we acquired a complicated tunneling term with an exponential electron-phonon
coupling  
$e^{\pm g(b^\dagger - b)}$. 

Although we have always used the original Hamiltonian in our
NRG calculations, the  transformed Hamiltonian of Eq.\ (\ref{eqn:h-prime})   
is very convenient to  gain some deeper insight into
the effective low-energy Hamiltonian generated by the renormalization
group transformations.This effective model is discussed in detail in Sec.\ \ref{sec:low-energy-h}
below.

\subsection{Numerical renormalization group and  the Anderson Holstein model}
\label{sec:nrg}

The properties of quantum impurity systems such  as defined
by $H=H_{imp} + H_{T} +H_{leads}$
can be very accurately calculated using the numerical renormalization group.
At the heart of this approach is
a logarithmic discretization of the continuous bath, controlled
by the discretization parameter 
$\Lambda > 1$.\cite{Wilson75,BullaCostiPruschke2008}
The continuum limit is recovered for $\Lambda \to 1$. Using
an appropriate unitary transformation,~\cite{Wilson75} the
Hamiltonian is mapped onto a semi-infinite chain, with the
impurity coupled to the first chain site. The $N$th link along the
chain represents an exponentially decreasing energy scale:
$D_N \sim \Lambda^{-N/2}$. Using this hierarchy of scales,
the sequence of finite-size Hamiltonians ${\cal H}_N$ for the
$N$-site chain is solved iteratively, discarding the high-energy states at
the conclusion of each step to maintain a manageable number
of states. The reduced basis set of ${\cal H}_N$ so obtained
is expected to faithfully describe the spectrum of the full
Hamiltonian on a scale of $D_N$, corresponding to the
temperature $T_N \sim D_N$. Details can be found on
the review\cite{BullaCostiPruschke2008} by Bulla et al..

Hewson and Meyer\cite{HewsonMeyer02} pioneered the 
application of the NRG on the single-lead 
version of Hamiltonian $H=H_0+ H_{ph}$. 
The standard NRG discretisation\cite{Wilson75,BullaCostiPruschke2008}
of  $H_{leads}$ maps the model onto a chain Hamiltonian
\begin{eqnarray}
\label{eqn:h-nrg}
H_N \Lambda^{-(N-1)/2} &=& 
 \tilde E_d d^\dagger d +  \tilde \w_0 b^\dagger b + \tilde \lambda_{ph} (b^\dagger + b)\left( d^\dagger d - \frac{1}{2}\right)
\non
&& 
+ \sum_{\alpha} V_\alpha (d^\dagger f_{0\alpha} + f_{0\alpha}^\dagger d) 
\non
&&
+ \sum_{\alpha}\sum_{n=0}^N  \Lambda^{-n/2} \bar \e_{n\alpha} 
f_{n\alpha}^\dagger f_{n\alpha}
\\
&&
+ \sum_{\alpha}\sum_{n=0}^{N-1}\Lambda^{-n/2} \bar t_n  (f_{n\alpha}^\dagger f_{n+1\alpha} + 
f_{n+1\alpha}^\dagger f_{n\alpha})
\nonumber
\end{eqnarray}
using a  NRG parameter $\Lambda>1$. $\bar t_n,\bar \e_n=O(1)$ and
$\bar \e_n=0$ for PH-symmetric leads.  The dimensionless parameters
$\tilde E_d,\tilde \omega_0,\tilde \lph$ are related to original
parameters\cite{Wilson75,BullaCostiPruschke2008} via the  scaling
factor $s=D(1+1/\Lambda)/2$: $\tilde E_d =  E_d/s, \tilde \w_0
=\w_0/s, \tilde\lambda_{ ph}=\lph/s$ and the renormalized tunneling
$V_\alpha = t_\alpha/s$. Using a suitable number of bosonic excitions
$N_b$,  $H_N$ is then iteratively diagonalized. A much more detailed
introduction to the NRG can be found in the Review by Bulla et
al..\cite{Wilson75,BullaCostiPruschke2008}

\subsection{Scattering-states numerical renormalization group  approach to quantum transport}
\label{sec:snrg}

For the calculating of the current $I(V)$ through the molecular level
at finite bias $V$ across the two leads, we have used the recently
proposed scattering-states numerical renormalization
group\cite{AndersSSnrg2008,SchmittAnders2010,SchmittAnders2011} (SNRG)
approach to quantum transport. The SNRG is based on an extension of
Wilson's NRG to non-equilibrium
dynamics.\cite{AndersSchiller2005,AndersSchiller2006,EidelsteinGuettgeSchillerAnders2012}
Using the  time-evolved density operator, the non-equilibrium steady
state retarded Green function can be calculated.\cite{AndersNeqGf2008}
Below, we only give a rather brief summary of the approach. A more
detailed derivation and discussion of the method can be found in
Refs.\
\onlinecite{AndersSSnrg2008,SchmittAnders2010,SchmittAnders2011}.

\subsubsection{Definition of the scattering states}
\label{sec:scattering-state}

In the absence of the electron-phonon interaction
the RLM defined in (\ref{eqn:two-lead-RLM})  
can be diagonalized exactly in the continuum limit\cite{JongHan2006,Hershfield1993,EnssSchoenhammer2005,HanHeary2007}
by  the following scattering-states creation operators
\cite{Oguri2007,LebanonSchillerAndersCB2003,AndersSSnrg2008,SchmittAnders2010} 
\begin{eqnarray}
  \label{eq:scattering-states-operators}
  \gamma^\dagger_{\e \alpha} &=& c^\dagger_{\e  \alpha} 
  + t_\alpha \sqrt{\rho_\alpha(\e)} G_{0}^r(\e+i\delta)
\\
&&
\hspace*{10mm}
 \times 
 \Bigg[
 d^\dagger
 +
 \sum_{\alpha'} 
 \int d\e' 
 \frac{t_{\alpha'} \sqrt{\rho_{\alpha'(\e')}}}{\e+i\delta -\e'}
 c^\dagger_{\e'\alpha'} 
 \Bigg]
 \punkt
  \nonumber 
\end{eqnarray}
$\alpha=L (R)$ labels left (right) moving scattering states created by
$\gamma^\dagger_{\e \sigma L(R)}$. In this equation, the local
retarded resonant-level Green function of the $d$-level  
\begin{eqnarray}
G^r_{0}(\w) &=& \left[\w+i\delta- E_d  - 
\Delta(\w+i\delta)
\right]^{-1}
\end{eqnarray}
enters as the expansion coefficient where $\delta>0$ is an
infinitesimally small energy scale required to select the correct boundary conditions.

Defining $\bar t = \sqrt{t_L^2 +t_R^2}$,  the ratio $r_{R(L)} = t_{R(L)}/\bar t $ 
denotes the relative tunneling strength of each lead $\alpha$ to the impurity $d$-level.
The resonant level self energy $\Delta_\alpha(\w+i\delta)$ in $G^r_{0}(\w)$
is given by
\begin{eqnarray}
\label{eq:DeltaDef}
\Delta(\w+i\delta) &=& \bar t^2 \sum_\alpha r^2_\alpha \int d\e \frac{\rho_\alpha(\e)}{\w+i\delta-\e}
\nonumber \\
& =& \Re e[\Delta(\w)] - i\Gamma(\w)
\end{eqnarray}
and its imaginary part $\Gamma(\w)$ denotes the energy dependent charge-fluctuation scale
which becomes a constant in the wide-band limes of a featureless band. 

In the limit of infinitely large leads
 the single-particle spectrum remains unaltered, and these scattering states diagonalize  the
Hamiltonian (\ref{eqn:two-lead-RLM}) 
\begin{eqnarray}
  \H_0 = \H(\lph=0) & =& \sum_{\alpha=L,R} \int d\e \, \e 
  \gamma^\dagger_{\e\alpha}\gamma_{\e\alpha}
  \label{eq:h0-U0}
\end{eqnarray}
up to a neglected ground state energy shift.\cite{Schweber1962}

The  creation operator $\gamma^\dagger_{\e \alpha}$  
is a solution of the  operator Lippmann-Schwinger equation\cite{Schweber1962}
and, therefore, the corresponding state break time-reversal
symmetry. The necessary  boundary condition for describing a
current-carrying open quantum system is encoded in the small imaginary part 
$+i\delta$ entering Eq.~(\ref{eq:scattering-states-operators}-\ref{eq:DeltaDef})
required for convergence when performing the continuum limit ${\rm
  Vol.}\to\infty$.  Since left and right movers at the same energy are
time-reversal pairs, time-reversal symmetry is restored at zero bias
yielding a exactly vanishing net current in that limit. 

To avoid  possible bound states, we will only consider the  wide-band
limit: $D\gg {\rm max}\{ | \e_d|,\Gamma_\alpha , |V|\}$. Furthermore,
$\Gamma_0=\Gamma(0) = \Gamma_L +\Gamma_R$ is used as energy unit in
this paper. We measure the coupling asymmetry by the ratio
$R=\Gamma_L/\Gamma_R$. A perfect unitary limit of $e^2/h$ conductance
quantum can only be reached for $R=1$ with  $R\to\infty$ and $R\to 0$
correspond to the tunneling regime. 

Hershfield  has shown that the steady-state density operator for a
current-carrying non-interacting  quantum system 
retains its Boltzmannian form\cite{Hershfield1993}
\begin{eqnarray}
  \hat \rho_0 &= &\frac{e^{-\beta(\H_0 -\hat Y_0)}}{\Tr{e^{-\beta(\H_0 -\hat Y_0)}}}
\, , \,
\non
 \hat Y_0 &=& \sum_{\alpha\sigma} \mu_\alpha \int d\e \,
 \gamma^\dagger_{\e\sigma\alpha}\gamma_{\e\sigma\alpha} 
 \label{eqn:rho_0}
\end{eqnarray}
at finite bias. The $\hat Y_0$ operator accounts for the
different occupation of the left-moving  and right-moving scattering states.
$\mu_\alpha$ denote the different chemical potentials of the leads.  
Since $H_0$ is bilinear, the transport is perfectly ballistic and the total current is given by the 
difference between the current to the left and the current to the right.

All steady-state expectation values of any operator are calculated 
using $\hat \rho_0$ for the non-interacting problem which includes the finite bias.  
In the absence of the electron-phonon
interaction this is a trivial and well-understood problem. 
It was  shown\cite{Oguri2007}  that the current obtained with this 
density-operator $\hat\rho_0$ is identical to the current calculated using a generalized Landauer
formula based on  Keldysh Green functions.\cite{HershfieldDaviesWilkins1991,MeirWingreen1992,MeirWingreen1994} 

This form of $\hat \rho_0$ stated in Eq.\ (\ref{eqn:rho_0}) 
remains valid even for the fully interacting system\cite{Hershfield1993} when replacing $\H_0\to \H$, 
and  replacing
$Y_0\to Y$. Since the $Y$ must be constructed from the many-body scattering states, its explicit analytical expression is unknown for a general Hamiltonian $H$ of an interacting systems.

We, therefore, proceed in two steps using the arguments outlined in the 
literature.\cite{Hershfield1993,DoyonAndrei2005}  
At first, we add to $H_0$ a fictitious electron-phonon interaction term $H_{e-ph}'$
which commutes with $Y_0$, i.e.\  $[H_{e-ph}', Y_0]=0$. Hershfield's argument
also yields a  steady-state density operator of the 
form $\hat \rho'_0= \exp[-\beta( \H_0 + H'_{e-ph} -\hat Y_0)]/Z$.
For the model under consideration we have chosen
\begin{eqnarray}
\hspace*{-5mm}
 H'_{e-ph} =
 \w_0 b^\dagger b + \lambda_{ph}  (b^\dagger + b)
 \left(\sum_{\alpha=L,R} r_\alpha^2 d^\dagger_\alpha d_\alpha - \frac{1}{2}\right)
\end{eqnarray}
where the $d_\alpha$ are the operators  obtained from inverting Eq.\ (\ref{eq:scattering-states-operators})
\begin{eqnarray}
d_\alpha &=& \bar t \int d\e \sqrt{\rho(\e)} [G^r_0(\e)]^* \gamma_{\e\alpha}
\, 
\end{eqnarray}
which  fulfill the anti-commutation relation $\{ d_\alpha, d^\dagger_\beta \} =\delta_{\alpha\beta}$.
The annihilation operator $d$ of a local electron on the device
is reconstructed by the linear combination of its left-mover and right-mover contributions  $d=r_R d_R + r_L d_L$.  We note that $H'_{e-ph}$ approaches $H_{e-ph}$ in the extreme tunneling
limit of $R\to \infty$ (or $R\to 0$.)

In a second step, we perform the time evolution of $\hat \rho'_0$ with respect to
the fully interacting Hamiltonian  to infinitely  long time:
the density operator $\hat \rho(t)$ progresses 
from its initial value $\hat \rho'_0$ at $t=0$ as
\begin{eqnarray}
\label{eqn:time-evolution-rho-0}
\hat \rho(t) &=& e^{-i\H_f t}\hat  \rho'_0 e^{i\H_f t}
\end{eqnarray}
where we set $\hbar=1$. 

\subsubsection{Scattering-states numerical renormalization group}

The basic idea of the  scattering-states numerical renormalization group  (SNRG) approach
is summarized as follows. 

(I) Knowing the analytical form of the non-equilibrium density operator $\hat \rho'_0$, we can discretize
scattering states on a logarithmic energy mesh identically as in the standard NRG\cite{BullaCostiPruschke2008,AndersSSnrg2008} and perform a standard 
NRG using $K_0=\H(\lph=0) + H'_{e-ph} -\hat Y_0$. The density operator $ \hat \rho'_0(V)$ contains
all information about the current carrying steady-state for the Hamiltonian $K_0+Y_0$.\cite{Hershfield1993}

(II) Starting at time $t=0$, we let the system evolve with
respect to the full Hamiltonian ${\H_f}=H(\lph>0)$  
Then, the density operator $\hat \rho(t)$ progresses 
from its initial value $\hat \rho'_0$ at $t=0$ according to Eq.\ (\ref{eqn:time-evolution-rho-0}).
Since we quench the system only locally, it is a fair assumption that
$\hat \rho(t)$ reaches a steady-state at $t\to\infty$ 
independent of initial condition for an infinitely large system, since
all bath correlation functions must decay for infinitely long times.

The finite size oscillations  always present in the NRG calculation \cite{AndersSSnrg2008} 
are projected out by defining the time-averaged  density operator
\begin{eqnarray}
\hat \rho_\infty &=& \lim_{\tau\to\infty} \frac{1}{\tau}\int_0^{\tau} dt \hat \rho(t)
\, .
\end{eqnarray}
As a consequence, only the matrix elements diagonal in energy contribute for $t\to \infty$ in accordance 
with the steady-state condition 
\begin{eqnarray}
[\H_f,\hat \rho_{\infty}] &=&0  \,\, .
\end{eqnarray}
Even though $\hat \rho_{\infty}$ remains unknown 
analytically, we can explicitly construct it numerically  using the 
time-dependent NRG\cite{AndersSchiller2005,AndersSchiller2006,EidelsteinGuettgeSchillerAnders2012}
(TD-NRG.)

(III) The steady-state retarded Green function is defined as
\begin{eqnarray} 
\label{eq:neqGf}  G^r_{A,B}(t) &= & - i \Tr{ \hat \rho_{\infty} [ \hat A(t), \hat B ]_s } \Theta(t), \label{eqn:steady-state-gf}   
\end{eqnarray}
where $\hat A(t)=e^{i\H_f t}\hat A e^{-i\H_f t}$, $[ \hat A(t), \hat B
]_s$ denotes the commutator ($s=-1$) for bosonic, and the
anti-commutator ($s=1$) for fermionic correlation functions. This
Green function can be calculated using the time-dependent
NRG\cite{AndersSchiller2005,AndersSchiller2006}  and extending ideas
developed for equilibrium Green
functions.\cite{PetersPruschkeAnders2006} The details of the algorithm
is derived in Ref.\ \onlinecite{AndersNeqGf2008}.

\subsubsection{Current as function of the bias voltage}

The current $I_\alpha$ is defined as a charge current\cite{MeirWingreen1992}
from the lead $\alpha$ to the local $d$-level.
It has been shown\cite{MeirWingreen1992,Hershfield1993,Oguri2007} that
for the model investigated here, the symmetrized  current
\begin{eqnarray}
  \label{eq:currentSym-def}
  I&=& r^2_R I_L-r^2_L I_R
  \quad .
\end{eqnarray}
is related to the steady-state spectral function of the local level, 
$\rho_d(\w,V) = \Im m [G^r_{\sigma}(\w-i0^+,V)]/\pi$ 
by a generalized Landauer formula 
\begin{eqnarray}
  \label{eq:ss-current}
  I(V) &=& \frac{G_0}{e}  \int_{-\infty}^\infty \, d\w \, 
\left[f_R(\w)-f_L(\w)\right] 
\Gamma_0 \pi \rho_d(\w,V)
\end{eqnarray}
where $f_\alpha(\w)=f(\w-\mu_\alpha)$.
The prefactor
\begin{eqnarray}
G_0 &=&   \frac{e^2}{h} \frac{4\Gamma_L
\Gamma_R}{\Gamma_0^2}
\label{eqn:G0-prefactor}
\end{eqnarray}
measures the asymmetry of the junction. 
$G_0$ reaches the universal conductance quantum $e^2/h$ for a symmetric 
point-contact junction, i.\ e.\ $\Gamma_L=\Gamma_R$, 
and is strongly suppressed in the tunneling regime $\Gamma_\alpha \ll \Gamma_{-\alpha}$.
The two chemical potentials  $\mu_\alpha$  are  set to $\mu_L=-r^2_R V$ and $\mu_R=
r^2_L V$ as function of the external source-drain voltage $V$
consistent with a serial-resistor model which is required by the  current conservation $I_R=-I_L$.

While  the conserving Keldysh Green function approach treats the
problem  of initially decoupled leads and propagate the system to a
current-carrying steady state, the  scattering states NRG starts from
an  initial current carrying steady state of the non-interacting
problem. Employing the
TD-NRG\cite{AndersSchiller2005,AndersSchiller2006} we let  the system
evolve by calculating the full density operator $\hat \rho(t)$
numerically  in the limit of $t\to\infty$ using the Hamiltonian of the
interacting system.  

In a continuum limit, any correlation function will have a finite
correlation time above which the memory of this initial state will be
lost. \cite{Hershfield1993, DoyonAndrei2005}  As a consequence, a
unique steady state  is approached which is compatible with  the
imposed boundary condition as pointed out by
Hersfield. \cite{Hershfield1993, DoyonAndrei2005} Our starting point
is therefore identical to any approach using  a propagation along a
Keldysh contour.\cite{Keldysh65}  The differences arise from the
inevitable  approximations made in Keldysh perturbation theory by
selecting a subclass of diagrams for a  practical calculation of the
non-equilibrium self-energies. In contrast, our approach does  not
make any of such approximations and, therefore, includes the full
interaction to infinitely high order. It only comprise a systematic
but well controlled error stemming from  discretization  of the bath
continuum\cite{BullaCostiPruschke2008,EidelsteinGuettgeSchillerAnders2012,GuettgeAndersSchiller2012}
inherent any NRG (or DMRG\cite{SchollwoeckDMRG2005}) approach. 

A hysteretic behavior\cite{LiHysteresis2003} is observed in some molecular junctions when sweeping the voltage a very slow but finite
rate. Using a simple mean-field approximation,  the possibility of
polaron formation on a molecular wire has been proposed as a mechanism for the observe NDR.\cite{GalperinRatnerNitzanHysteresis2004}
Such hysteretic behavior in  self-consistent equation can also occur
in the employed approximations  when increasing the coupling constant
beyond the validity range of such approximations. Such many-valued
solution might not survive in an exact solution of the problem and
their robustness need to be checked against the inclusion of higher
order contributions. A prominent example of such a known false
hysteresis in the I-V characteristics is the GW-approach to Anderson
model out of equilibrium.\cite{spataruGwMillis2009} 

The theoretical description of such hysteretic behavior requires tracing the same experimental conditions in the simulations. Starting from 
the  initial condition of a fully interacting current-carrying steady
state for a given bias voltage  the time dependent current
$I(V(t),t)$ must be calculated for a given rate of voltage change.

In a recent paper by Muehlbacher et al.\cite{MuehlbacherThoss2012} 
it has been proposed that signatures of bistability  
in non-equilibrium charge transport through molecular quantum dots
might be linked to subtle differences in the initial conditions.
This, however,  is clearly beyond our scattering-states approach which targets
exclusively the stationary  steady-state limit.

\section{Equilibrium renormalization group approach}
\label{sec:equi-nrg}

Before we present the results for the quantum transport in 
Sec.\ \ref{sec:quantum-transport},
we briefly review the equilibrium parameter flow of the model. 
This provides the  necessary understanding of the energy scales 
and defines the different regime which become relevant for the
quantum transport.

\subsection{Low energy Hamiltonian}
\label{sec:low-energy-h}

While Hewson and Meyer investigated
the equilibrium dynamics of the more complex 
spin-full model\cite{HewsonMeyer02} 
using the NRG, we focus on the simpler spin-less model  in this work.
Its  low energy fixed point is also given  by an effective resonant level model. 
In the limit  $\w_0>D,\Gamma_0$, its effective
tunneling matrix elements $t^{\rm eff}_\alpha$ are  estimated by using $H'$ defined in 
Eqn.\ (\ref{eqn:h-prime}):
the local phonons are approximately in their ground state, and 
\begin{eqnarray}
t_\alpha\to t^{\rm eff}_{\alpha}&\approx& t_\alpha\bra{0}  \exp(-g(b^\dagger - b))\ket{0} \non
&=& t_\alpha \exp(-\frac{g^2 }{2})
\punkt
\end{eqnarray}
In leading order,  we obtain the renormalized charge fluctuation scale $\Gamma_{\rm eff} \approx\Gamma_0 \exp( -g^2)$ in this limit.

By expanding the factor
\begin{eqnarray}
\label{eqn:tunneling-expansion}
e^{-g(b^\dagger - b)} &=&
1 -g(b^\dagger - b)
+O\left(g^2\right)
\end{eqnarray}
for small $g\ll 1$ added by 
the Lang-Firsov transformation to  the tunneling term
\begin{eqnarray}
H'_{T} &=& \sum_\alpha t_\alpha (e^{g(b^\dagger - b)} \bar d^\dagger c_{0\alpha} + c^\dagger_{0\alpha} \bar d e^{-g(b^\dagger - b)})
\,\, ,
\end{eqnarray}
it is apparent that an effective repulsive Coulomb interaction term 
\begin{eqnarray}
H_{U} &=& \sum_\alpha U_{\rm eff} (d^\dagger d - \frac{1}{2}) ( n_{0\alpha}-\frac{1}{2})
\end{eqnarray}
is obtained in second-order perturbation theory in $g$. 
Of course, this type of effective interaction and all possible higher order contribution 
will be automatically generated by  RG transformation in each NRG iteration step.
Taking into account the renormalization of the tunneling matrix element 
$t_\alpha\to t^{\rm eff}_{\alpha}$, we conjecture that
$U_{\rm eff}$ generated in the RG transformation is given by 
\begin{eqnarray}
U_{\rm eff} &\propto &  \left(\sum_\alpha t_{\alpha}^{\rm eff} g \right)^2 \frac{1}{\w_0} 
\end{eqnarray}
in leading order. Then, the dimensionless  Coulomb repulsion reads
\begin{eqnarray}
\frac{U_{\rm eff}}{D} &=&   
\frac{\pi}{2} \frac{\Gamma_{\rm eff}}{\w_0}g^2
f(x)
=\frac{U_{ana}}{D} f(x)
\label{eqn:U-eff-ana}
\end{eqnarray}
where $f(x)$ is an unknown scaling function of order $O(1)$ which accounts for higher order corrections
in the expansion (\ref{eqn:tunneling-expansion}). The analysis of the NRG data, presented below
suggests  that the scaling variable  is given by
$x=\alpha(\w_0/\Gamma_0) g^2$ in the anti-adiabatic regime $\w_0\gg \Gamma_0$.
The function $\alpha(y)$ is a slow varying function of the order $O(1)$  as shown
in Sec.\ \ref{sec:NRG-renormalized-params}.

A perturbative treatment\cite{EidelsteinSchiller2012} to second order in $t_\alpha$, predicts
$U_{\rm eff}/D\approx (4\Gamma_0 E_{p})/(\pi \w_0^2)$ in the weak coupling limit, 
i.\ e.\ $g\ll 1$, and  $U_{\rm eff}/D\approx (4\Gamma_0/E_{p})$ in the strong coupling limit, 
$g\gg 1$.
Note that these results are only valid as long as $\Gamma_0$  is the smallest energy scale
and, therefore, $U_{\rm eff}/D$ remains a perturbative correction in the validity range 
of the perturbation theory. This analytic result
agrees nicely with our conjecture of Eq.~(\ref{eqn:U-eff-ana}) which will be backed
up by our extensive NRG study below. 

Note, however, that the values for 
$U_{\rm eff}/D$ extracted from  the full NRG calculation can result in 
magnitudes of $O(1)$  and, therefore, 
this regime would be beyond the reach of a 
second-order perturbation theory. Nevertheless, the leading-order scaling with the model
parameters remains well captured by  Eq.~(\ref{eqn:U-eff-ana})  as long as $\w_0/\Gamma_0>1$.

\subsection{Renormalized parameters}
\label{sec:renormalized-parameters}

In equilibrium,  the anti-binding combination of lead operators, $c_{0-}= r_L c_{0R} -r_R c_{0L}$
decouples from the impurity, and we are left with an effective one-band model since the
local $d$-orbital is only connected to orthogonal combination $c_{0+}= r_R c_{0R} +r_L c_{0L}$.
Therefore, we focus  on an effective one-band model but 
with the charge-transfer rate $\Gamma_0+\Gamma_L+\Gamma_R$.

Eidelstein et al.\cite{EidelsteinSchiller2012}
have already pointed out that the low-energy Hamiltonian of
the spinless one lead 
Anderson Holstein model can be mapped onto an effective interacting resonant level model (IRLM)
\begin{eqnarray}
\label{eqn:IRLM-def}
H^{\rm IRLM} &=& H_0 + H_{imp}
\non
H_0 &=& \sum_{k} \e_{k}c^\dagger_{k}c_{k}
\non
&&
+
 t_{\rm eff}  (d^\dagger c_{0} + c^\dagger_{0} d)
\\
H_{imp} &=& \tilde \epsilon_d d^\dagger d +U_{\rm eff} (d^\dagger d -\frac{1}{2})
(c^\dagger_{0} c_{0} -\frac{1}{2})
\nonumber
\end{eqnarray}
for effective band width $D_{\rm eff} <\w_0$. 
Its parameters $\tilde \e_d,t_{\rm eff}$ and $U_{\rm eff}$ can be extracted
in different ways from the NRG level spectrum.\cite{Wilson75,HewsonMeyer02,BullaCostiPruschke2008,EidelsteinSchiller2012}

In Ref.\ \onlinecite{EidelsteinSchiller2012}, the effective parameter
$\Gamma_{\rm eff}\propto t^2_{\rm eff}$ is extracted from the charge
susceptibilty at the particle-hole symmetric point. We, however, use a
different approach which  allows for a definition of  $\Gamma_{\rm
  eff}$ for arbitrary values of $E_d$,  away from the particle-hole
symmetric point by applying   the procedure proposed by Hewson et
al..\cite{HewsonOguriMeyer2004}  

In this method the renormalized parameters are extracted directly from
the NRG level flow. It is  based on the analytically known low-energy
stable Fermi-liquid fixed point\cite{BullaCostiPruschke2008} and the
exact solution of the RLM for a given NRG chain. The  discretized
effective RLM for the fixed-point dynamics, 
\begin{eqnarray}
\label{eqn:h-nrg-i}
 H_N^{RLM} \Lambda^{-(N-1)/2} & =&
 \tilde \e_d d^\dagger d 
 + t_{\rm eff}(d^\dagger f_{0} + f_{0}^\dagger d) 
\non
&&
+ \sum_{n=0}^N  \Lambda^{-n/2} \bar \e_{n} 
f_{n}^\dagger f_{n}
\\
&&+ \sum_{n=0}^{N-1}\Lambda^{-n/2} \bar t_n  (f_{n}^\dagger f_{n+1} + 
f_{n+1}^\dagger f_{n})
\nonumber
\, ,
\end{eqnarray}
is exactly diagonalized\cite{Wilson75,BullaCostiPruschke2008} 
to
\begin{eqnarray}
\label{eqn:LT-fixed-point}
H_N^{RLM} &=&
\sum_{l=1}^{\frac{N+2}{2}}
\left(\e_{p,l} e^\dagger_{l} e_l 
+
\e_{h,l} h^\dagger_{l} h_l 
\right)
\end{eqnarray}
for a NRG chain with odd number of lead sites -- $N$ even. 
$e^\dagger_{l}$ ($h^\dagger_{l}$) creates the $l$-th  elementary particle (hole) excitation
with the energy $\e_{p,l}$  ($\e_{h,l}$.)
Obviously, the impurity Green function $G^N_d(z)$ 
\begin{eqnarray}
G^N_d(z) &=& \frac{1}{ z -\tilde \e_d\Lambda^{(N-1)/2} - [V_{\rm eff}]^2\Lambda^{N-1} g_{0}(z)}
\end{eqnarray}
must have poles at these single-particle excitation energies $z=\e_{e,l},\e_{h,l}$. $g_0(z)$ is the 
Green function at chain site $n=0$ in the absence of the impurity and can be calculated by a
continuous fraction expansion.\cite{HewsonOguriMeyer2004} 

The first single-particle excitation, $E_{1p}$, and the first
single-hole excitation of the NRG level spectrum,  $E_{1h}$
respectively, are very good approximation  of $\e_{p,1}$ and
$\e_{h,l}$ close to the Fermi-liquid fixed point. These two NRG
energies are sufficient to determine  the two unknown effective
parameters $\tilde \e_d$ and $V_{\rm eff}$ by solving the two coupled
algebraic equations\cite{HewsonOguriMeyer2004} 
\begin{eqnarray}
 \frac{E_{1p} \Lambda^{-(N-1)/2}}{2\Gamma_{\rm eff}(N)} -\frac{\tilde \e_d(N)}{2\Gamma_{\rm eff}(N)} 
&=& \frac{\Lambda^{(N-1)/2}}{ \pi} g_0(E_{1p})
\\
 \frac{-E_{1h} \Lambda^{-(N-1)/2}}{2\Gamma_{\rm eff}(N)} -\frac{\tilde \e_d(N)}{2\Gamma_{\rm eff}(N)} 
&=& \frac{\Lambda^{(N-1)/2}}{ \pi} g_0(-E_{1h})
\nonumber
\end{eqnarray}
for the poles of the Green function, where $\Gamma_{\rm eff}=\pi
t_{\rm eff}^2\rho(0)$.  Note, that these are dimensionless parameters,
all given in units of $D(1+ 1/\Lambda)/2$. 

In order to calculate $U_{\rm eff}$, we analyze the four eigenstates
of  $H_0^{{\rm IRLM}}$ comprising only  the local level and $f_0$. It
is easy to see that $U_{\rm eff}$ is related to the energy difference
between the lowest particle-hole excitation and the sum of a
single-particle  and a single-hole excitation. The adaptation of
equation (18) in Ref.\ [\onlinecite{HewsonOguriMeyer2004}] 
\begin{eqnarray}
\label{eqn:U-eff}
&&E_{1p}+E_{1h} - E_{1ph} = 2U_{\rm eff}(N) \Lambda^{(N-1)/2}\non
&&\times \left(
|\psi_{d,p}(1)|^2 |\psi_{c,h}(1)|^2 +
|\psi_{d,h}(1)|^2 |\psi_{c,p}(1)|^2
\right)
 \, \,
\end{eqnarray}
requires the knowledge of expansion coefficients $\psi_{d,p}(l)$
of the annihilation operator $d$
\begin{eqnarray}
d &=& \sum_{l=1}^{(N+2)/2} \left( \psi_{d,p}(l) e_{p,l} + \psi_{d,h}(l) h^\dagger_{h,l}\right)
\end{eqnarray}
and  $\psi_{c,p}(l)$ of the first lead chain site $n=0$
\begin{eqnarray}
f_0 &=& \sum_{l=1}^{(N+2)/2} \left( \psi_{0,p}(l) e_{p,l} + \psi_{0,h}(l) h^\dagger_{h,l}\right) 
\end{eqnarray}
in terms of the elementary excitation of the low-energy fixed point Hamiltonian (\ref{eqn:LT-fixed-point}).
The coefficients $|\psi_{d,p}(l)|^2$  are determined by the weights
of the poles of the finite size chain impurity Green functions
\begin{eqnarray}
G^N_d(z) &=& \sum_{l=1}^{(N+2)/2}\left(
 \frac{|\psi_{d,p}(l)|^2}{z-e_{p,l}}
 +
  \frac{|\psi_{d,h}(l)|^2}{z-e_{h,l}}
  \right)
\end{eqnarray}
and  coefficients  $|\psi_{c,h}(l)|^2$ by the corresponding
Green functions of the first lead site $n=0$, $G^N_c(z)$, which is related to the local
$T$-matrix $V^2G_d^N(z)$ 
\begin{eqnarray}
G^N_c(z)&=& g_0(z) + V_{\rm eff}\Lambda^{N-1} G^N_d(z) g^2_0(z)
\\
&=& g_0(z) \left[
\frac{z-\tilde \e_d^{\rm eff}\Lambda^{(N-1)/2} }{ z -\tilde \e_d^{\rm eff}\Lambda^{(N-1)/2} - [V_{\rm eff}]^2\Lambda^{N-1} g_{0}(z)}
\right]
\nonumber
\end{eqnarray}
Using the calculated parameters  $\tilde \e_d(N)$ and $\Gamma^{\rm eff}(N)$, the
spectral weights at  the pole $\e_{p,1}$ are approximately given by
\begin{eqnarray}
|\psi_{d,p}(1)|^2 &=& \frac{1}{1- [V_{\rm eff}]^2\Lambda^{N-1} g'_{0}(E_{1p})} \non
|\psi_{c,p}(1)|^2 &=& g_0(E_{1p})\frac{E_{1p} -\tilde \e_d\Lambda^{(N-1)/2} }{1- [V_{\rm eff}]^2\Lambda^{N-1} g'_{0}(E_{1p})} 
\end{eqnarray}
and analog for $\e_{h,1}$. We have replaced the true
single-particle (single-hole) excitation energy
$\e_{p,1}$ ($\e_{h,l}$)
by the first NRG particle (hole) excitation energy $E_{1p}$ ($E_{1h}$.)
Thereby, $g'_{0}(z)$ denotes the derivative of $g_0(z)$.

\subsection{NRG analysis of the renormalized parameter}
\label{sec:NRG-renormalized-params}

\begin{figure}[htbp]
\begin{center}
\includegraphics[width=85mm]{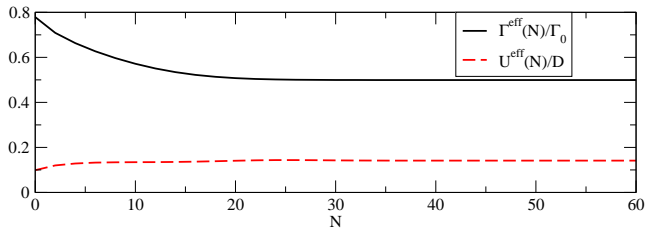}

\caption{(Color online.) Example NRG flow for $U(N)/D$ and $\Gamma_{\rm eff}(N)$
vs NRG iteration $N$ for the parameters
$\w_0/\Gamma_0= \lph/\Gamma_0=5$.
}
\label{fig:1}
\end{center}
\end{figure}

In this section, we present the results for the renormalized parameter
for the low-energy equilibrium Hamiltonian extracted from the NRG
level spectrum  of the single-lead model as described in section
\ref{sec:renormalized-parameters}.  We have used $\Lambda=1.5$, kept
$N_s=1500$ NRG states after each iteration, and included the lowest
$N_b=400$ local phonon states. The band-width of a lead with constant
DOS $\rho(\w)=\Theta(D-|\w|)/2D$ was set to $D=100\Gamma_0$, and
$\Gamma_0$ serves as energy  scale throughout the paper. In this
section, we only investigate the PH-symmetric limit by setting
$E_d=0$. 

\begin{figure}[tb]
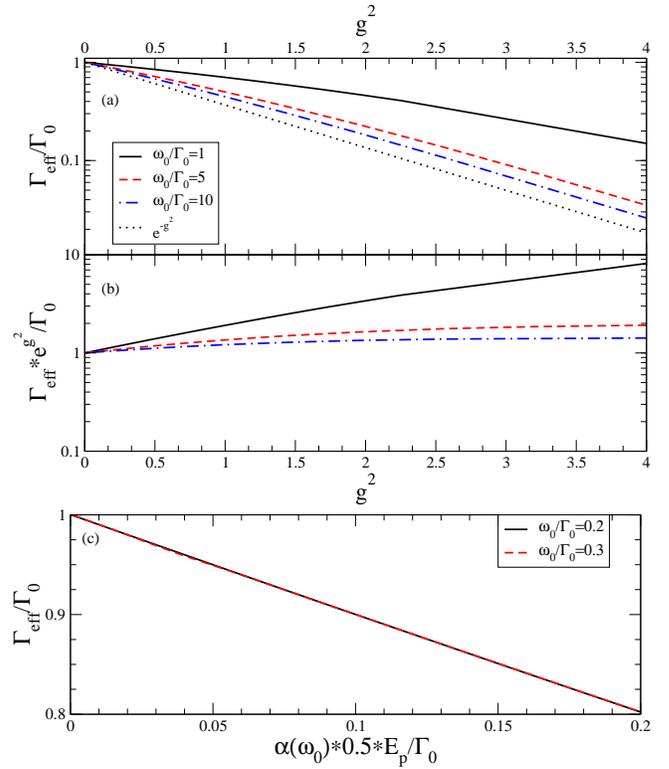

\begin{center}
\includegraphics[width=85mm]{fig-3-ab-gamma-eff-nrg.eps}

\includegraphics[width=85mm]{fig-3-c-gamma-eff-adiabatic.eps}

\caption{(Color online.)  (a) Anti-adiabatic and crossover regime: $\Gamma_{\rm eff}$ as function of 
$g^2=(\lph/\w_0)^2$ for
$\w_0/\Gamma_0=1,5,10$, (b) rescaled  $\Gamma_{\rm eff} e^{g^2}$
vs $g^2$;
(c) $\Gamma_{\rm eff}/\Gamma_0$ vs $E_{\rm p}$ in the adiabatic regime. The correction factor $\alpha(\w_0=0.2)=1.02,\alpha(\w_0=0.3)=0.93$. 
In the anti-adiabatic and the crossover regime, (a)  and (b) $N_s=1500$, 
in the adiabatic regime  (c) we kept $N_s=2500$ NRG states
for accuracy.
}
\label{fig:2}
\end{center}
\end{figure}

In Fig.\ \ref{fig:1}, the flow of $\Gamma_{\rm eff}(N)$ and $U_{\rm
  eff}(N)$ is depicted with respect to the NRG iteration $N$ for one
specific parameter set, $\w_0/\Gamma_0= \lph/\Gamma_0=5$.  Fast convergence
\footnote{Since the fixed point of the model is a non-interacting RLM
  and, therefore, energy difference vanishes for $N\to\infty$ on the
  l.h.s.\ of Eq.\ (\ref{eqn:U-eff}), the extracting of $U_{\rm eff}$ becomes numerically unstable once the difference 
is smaller than $10^{-12}$, and we need to stop the procedure.}
is achieved when approaching the low-energy fixed point for
$N\to\infty$.  We have  extracted the fixed point values of  
$ \e^{\rm eff}_d=\lim_{N\to\infty} \tilde \e_d^{\rm eff}(N)$,
$ \Gamma_{\rm eff}=\lim_{N\to\infty}  \Gamma_{\rm eff}(N)$,
$ U_{\rm eff}=\lim_{N\to\infty} U_{\rm eff} (N)$ for various values of the electron-phonon coupling 
$\lambda_{ph}$ and $\w_0$.
Since, $ \e^{\rm eff}_d=0$ for all PH-symmetric parameters, we do not show any results. In the ph-asymmetric regime, $ \e^{\rm eff}_d\neq0$ measures the effective PH-symmetry breaking external field.

The results for $\Gamma_{\rm eff}/\Gamma_0$ vs $g^2$ are shown in
Fig.\ \ref{fig:2}(a) for three different phonon frequencies
$\w_0/\Gamma_0=1,5,10$. For $\Gamma_0\ll \w_0$, we approach the
anti-adiabatic limit in which the phonons can almost instantaneously
react on the charge fluctuations. The data suggests that  $\Gamma_{\rm
  eff} \approx \Gamma_0\exp[-g^2]$ for $\w_0\to\infty$ and weak and
moderate values of $g$. The closer $\w_0$ approaches the
charge-fluctuation scale $\Gamma_0$, the stronger the deviations for
this approximation. In the context of the spin-full model,  Hewson et
al.\ \cite{HewsonMeyer02} have already  proposed a modification of the
exponent $\exp(-g^2)\to \exp(-g^2 h (\lph,\w_0))$ by a function $h$
which accounts for the additional correction. For large $\w_0$, the
leading contribution to $h$ is a constant approaching $1$, however, we
were not able to find a universal scaling function $h$ using our data
which describes the crossover from weak  to strong coupling,   $g <
1$  and $g >  1$ respectively .

Restricting to the  strong coupling limit, however, Eidelstein et
al. \cite{EidelsteinSchiller2012} were able to show that all graphs
$\Gamma_{\rm eff}(g,\w_0)$ collapse onto a single curve by
introducing the  universal function through the expression
$\Gamma_{\rm eff}/\Gamma = \exp[- g^2 F(R)]$.  The dimensionless
parameter $R=\Gamma_0/E_{p}$ emphasizes  the importance of the
polaronic energy shift $E_{p}$ for the physics of the model.  

The validity range of such a universal function $F(R)$ requires $R\ll 1$. 
For $R\to 1$, however, significant deviations from a universal scaling are already observed in 
Ref.\ \onlinecite{EidelsteinSchiller2012}. 

Since we focus on the regime $R>1$,
the study by Eidelstein et al. \cite{EidelsteinSchiller2012} investigates the complementary
strong-coupling regime to our weak and intermediate coupling regime. Although the equilibrium NRG
can reach $R\ll 1$ as recently demonstrated, the TD-NRG and the SNRG are restricted to moderate
values of $g$
due to the increasing of discretization artefacts.\cite{EidelsteinGuettgeSchillerAnders2012,GuettgeAndersSchiller2012}

In the adiabatic regime, $\w_0\ll \Gamma_0$, field-theoretical arguments\cite{VinklerSchillerAndrei2012} have been employed to show
\begin{eqnarray}
\label{eqn:adiabatic-gamma-eff}
\Gamma_{\rm eff} &\approx &  \Gamma_0
\left( 1 -\frac{2}{\pi}\frac{E_{\rm p}}{\Gamma_0}
\right)
\end{eqnarray}
for $ {max}\{E_{\rm p}/\Gamma_0, g,E_d/\Gamma_0\} \ll 1$. Our
numerical analysis for the adiabatic regime agrees with that
perturbative result, but we find a prefactor of $1/2$ up to very small
corrections instead of $2/\pi$ stated in  Eq.\
(\ref{eqn:adiabatic-gamma-eff}), as shown in Fig.\ \ref{fig:2}(c). In
the analytical calculation\cite{VinklerSchillerAndrei2012} the local
Lorentzian spectral function was approximated by a simple constant
$\rho$ while in our numerics the full energy dependency enters the
calculations which probably accounts for the difference. Our data
provides evidence for a crossover  from a perturbative $[ 1
-\alpha\frac{E_p}{\Gamma_0}]$  correction factor to $\Gamma_0$ with
$\alpha\approx 1/2$ in the adiabatic regime  to an exponential
suppression factor $\exp[ - g^2h(\lph,\w_0)]$ in the anti-adiabatic regime.

Let us now discuss $U_{\rm eff}$ dynamically generated by the
electron-phonon interaction.   The plot in Fig.\ \ref{fig:1}
demonstrates clearly that  such an effective Coulomb repulsion $U(N)$
can be extracted directly  from the NRG level flow via Eq.\
(\ref{eqn:U-eff})  and converges rather quickly to a low temperature
fixed point value $U_{\rm eff}$.

\begin{figure}[tbp]
\begin{center}
\includegraphics[width=85mm]{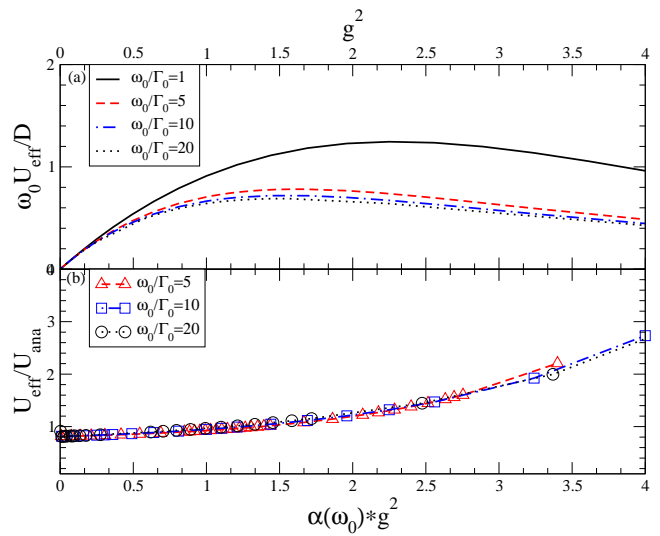}

\caption{(Color online.)  (a) $\w_0 U_{\rm eff}$ as function of $g^2$ for four different values of
$\w_0/\Gamma_0=1,5,10,20$.
(b)  $f(x)=U_{\rm eff}/U_{ana}$ as defined in Eq.\ (\ref{eqn:U-eff-ana}) vs 
$\alpha(\w_0)g^2$, where we used
the scaling factor $\alpha(\w_0)=0.85,1,1.1$ for $\w_0/\Gamma_0= 5,10,20$.
NRG parameters as in Fig.\ \ref{fig:2}.
}
\label{fig:3-U-eff}
\end{center}
\end{figure}

The results for $\w_0\, U_{\rm eff}$ are depicted in Fig.\
\ref{fig:3-U-eff}(a) for four different values of
$\w_0$. Qualitatively, the results can be understood by the
conjectured form 
\begin{eqnarray}
\frac{U_{\rm ana}}{D} &=&   
\frac{\pi}{2} \frac{\Gamma_{\rm eff}}{\w_0} g^2 
=
\frac{\pi}{2} \frac{\Gamma_{\rm eff} E_p}{\w^2_0}
\, .
\nonumber
\end{eqnarray}
 of Eq.\ (\ref{eqn:U-eff-ana}) above where $U_{\rm eff}\propto U_{\rm ana}$.
 $\w_0 U_{\rm eff}$ increases linearly with $x=g^2$ and is exponentially suppressed
for large $x$ due to the renormalization of the bare hopping constant. Focusing only on the three
graphs in the anti-adiabatic regime $\w_0 >\Gamma_0$, we plot the ratio $U_{\rm eff}/U_{\rm ana}$
as function of $\alpha(\w_0)x$.

By an appropriate dimensionless scale $\alpha(\w_0)$ which only
depends on the phonon frequency, we are able to rescale $x$ such that
all ratios $U_{\rm eff}/ U_{ana}$  collapse onto one universal scaling
curve of the order $O(1)$ in this regime as shown in Fig.\  \ref{fig:3-U-eff}(b).

The simple arguments  leading to the estimated analytical value
$U_{ana}$ do not hold  in the extended anti-adiabatic  and in the
crossover regime,  and this  mapping onto the universal scaling curve
fails for $\w_0/\Gamma_0=1$.  As indicated already in  Fig.\
\ref{fig:3-U-eff}(a),  $U_{\rm eff}$ is much larger and the additional
renormalization for $\Gamma_{\rm eff}$ by the finite $U_{\rm eff}$  in
the effective IRLM might have to be taken into account when crossing
over to the adiabatic regime $\w_0<\Gamma_0$.

\section{Equilibrium spectral functions}
\label{sec:equilibrium-rho}

\subsection{Technical details}

Since the equilibrium spectral functions are used to calculate the
zero-bias conductance within the linear response theory, we present
some local spectra to set the stage for the non-equilibrium transport.
Furthermore, we also can use the direct calculation of the equilibrium
spectral function as a benchmark for testing the quality of the
non-equilibrium spectra obtained by the time-dependent NRG
approach\cite{AndersSchiller2005,AndersSchiller2006} to
non-equilibrium Green  functions.\cite{AndersNeqGf2008} 

For two trivial limits, exact solutions are analytically known.
 (i)
In the absence of the coupling to the leads, $t_\alpha=0$,  the exact
solution of the local spectral function\cite{Mahan81} comprises of a
set of equidistant $\delta$-peaks separated by the phonon-frequency
$\w_0$.  Their spectral weights are given  by modified Bessel
functions  whose arguments are temperature dependent.\cite{Mahan81}  
(ii) 
In the opposite limit,  $\lambda_{ph}=0$,  the  spectral function of is simply given by
the Lorentzian of the RLM which will be weakly modified for $0<g\ll 1$.

We have used the complete Fock space
algorithm\cite{PetersPruschkeAnders2006,AndersNeqGf2008,WeichselbaumDelft2007}
to calculate the NRG spectral function. The $\delta$-functions of the
spectral function have been  replaced by the  standard NRG broadening
function  
\begin{eqnarray}
\delta(\w\pm |\w_n|)&\to& \frac{\exp(-b^2/4)}{b|\w_n|\sqrt{\pi}} e^{ -[\frac{1}{b}\log(|\w/\w_n|]^2}
\end{eqnarray}
characterized by the broadening parameter  $b$ and additionally
averaged over $N_z$ different band
discretizations.\cite{YoshidaWithakerOliveira1990,AndersSchiller2005,AndersSchiller2006}
The equation of motion\cite{BullaHewsonPruschke98} exactly relates the
self-energy  
\begin{eqnarray} 
\label{eq:nrg-self-energy}
\Sigma^{\rm ph}(z) &=&  \lph \frac{F(z)}
{G_d(z)}
\end{eqnarray}
to ratio of the correlation function
\begin{eqnarray}
F(z)=\correl{(b+b^\dagger) d}{d^\dagger}
\end{eqnarray}
and the local
Green function $G_d(z)=\correl{ d}{d^\dagger}$.  

For $\lph=0$, the correlation function $F(z)$ vanishes: it  is
generated in first order by the electron phonon coupling. Therefore,
$F(z)\propto \lph$ in leading order which immediately reproduces the
leading order of magnitude of $\Sigma^{\rm ph}(z) \propto \lph^2$ for
the weak coupling regime $\lph/\Gamma_0 \ll 1$ obtained from
perturbation theory\cite{Caroli71} where the leading order Feynman
diagram is  depicted in Fig.~\ref{fig:self-electron-phonon-diagram}.

The spectral functions discussed below are all obtained from $G_d(z)=
[z-E_d -\Delta(z) - \Sigma^{\rm ph}(z)]^{-1}$ where the total
self-energy is given by the sum of $\Sigma^{\rm ph}(z)$ and
$\Delta(z)$  defined in Eq.\ (\ref{eq:DeltaDef}). In our numerics we
evaluate  the Green function very close to the real axis  and
$\rho_d(\w)= \Im m G_d(\w-i\delta)/\pi$   using a very small value for
$\delta/\Gamma_0 =10^{-10}-10^{-7}$.

\subsection{Particle-hole symmetry}

\subsubsection{Crossover regime}

\begin{figure}[tbp]
\begin{center}
\includegraphics[width=85mm]{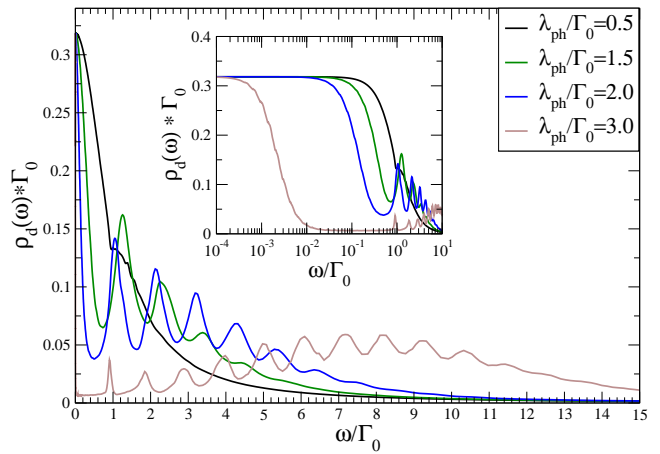}

\caption{
(Color online.) 
Equilibrium spectral functions for $E_d=0$, $\w_0/\Gamma_0=1$ and various values of 
$\lph/\Gamma_0=0.5,1.5,2,3$.
%
NRG parameter: $N_s=1200$, $\Lambda=1.5$, $N_b=80$, $N_z=512$, $b=0.03$.
}
\label{fig:gf-specta-crossover-i}
\end{center}
\end{figure}

In Fig.\ \ref{fig:gf-specta-crossover-i}, the positive part of the
symmetric local spectral function $\rho_d(\w)=\Im m
G_d(\w-i\delta)/\pi$ is depicted for a series of electron-phonon
couplings  $\lph/\Gamma_0=0.5,1.5,2.0,3.0$ and a fixed phonon
frequency $\w_0/\Gamma_0=1$.  For $\lph/\Gamma_0=0.5$, the RLM
Lorentzian is broadened and a kink at $\w\approx \w_0$ is related to a
sharp rise of the self-energy at $\w$. With increasing $\lph$, more
and more phonon replica of the resonance peak $\w=E_d=0$ are visible
and their width increasingly narrows which is related to the narrowing
of the central peak as depicted in the inset.  Obviously the central
peak width is given by the low temperature fixed point value
$\Gamma_{\rm eff}$ which  is already exponentially suppressed for
$\lph/\Gamma_0=3$.  

We have used a very large number of $N_z=512$ values  in combination
with a very small broadening $b=0.03$ for the $z$-averaging of the
spectral functions to ensure that the spectra depicted in Fig.\
\ref{fig:gf-specta-crossover-i} are really nearly broadening
independent even at higher frequencies.

The spectral peak at $\w=0$ remains pinned at its universal value for
$T=0$ in accordance  with the Friedel
sum-rule.\cite{Langreth1966,Anders1991,EidelsteinSchiller2012}
The spectral weight of this zero-frequency peak, however, is reduced
to $\Gamma_{\rm eff}/\Gamma_0$ and redistributed to the higher energy phonon replica peaks.

For $\lph/\Gamma_0=3$, we checked in  several very  long NRG runs
averaging up to  $N_z=1024$ bath discretizations and using a very
small broadening parameter of $b=0.01$  that   the line width of the
high energy phonon peaks are  independent of the NRG broadening
selected in Fig. \ref{fig:gf-specta-crossover-i}. 

The increase of the spectral function and the maxima of the envelop
functions at $\w\approx \pm 8\Gamma_0$ is related to the  finite
$U_{\rm eff}$.  Considering only the two local orbitals, $d$ and
$c_0$, of the IRLM  (\ref{eqn:IRLM-def}) which defines the initial NRG
Hamiltonian $H_0^{NRG}$ of the IRLM model, one can calculate the
spectral functions exactly and finds  single-particle excitations at
$\w_N\approx \pm U(N)/2$. Since the value of $U(N)$ can strongly
depend on the iteration and approaches its fixed point value $U_{\rm
  eff}$  only for $|\w|\ll \Gamma^{\rm eff}$, the peak positions
correspond to the value of the renormalized $U(N)/2$ where the NRG
iteration $N$ corresponds to the energy scale
$\w\approx\w_N\propto\Lambda^{-(N-1)/2}$. This is in complete analogy
to the single-impurity  Anderson model where the bare high-energy
value of $U$ defines the location of the high-energy charge excitation
peaks, while the low-energy  fixed point value of $U_{\rm siam}^{\rm
  eff} \approx T_K$ as shown by Hewson et
al.\cite{HewsonOguriMeyer2004}

Although Fig.\ \ref{fig:3-U-eff}(a) indicates that for
$\lph/\Gamma_0=3$, $U_{\rm eff}$ is significantly lower than for the
smaller values $\lph/\Gamma_0=1.5,2$, no additional charge peaks are
found for the latter parameters. This is related to the fact that the
self-energy contribution $\Sigma^{\rm ph}(z)$ is proportional to
$\lph^2$ in leading order. If the magnitude of the self-energy is too
small, the  renormalized resonant level width $\Gamma(N)$ dominates
the envelop function. Once $g>1$, $\Gamma_0$ evidently becomes
significantly suppressed, and spectral weight is increasingly shifted
to higher frequencies. The broad phonon side peaks become observable
once $\Gamma_{\rm eff} \ll \Gamma_0$ and the central resonance lost
much of its weight.

\subsubsection{Weak coupling regime}
\label{sec:weak-coupling}

\def\fmfL(#1,#2,#3)#4{\put(#1,#2){\makebox(0,0)[#3]{#4}}}

\begin{figure}[t]

\begin{picture}(200,100)(0,0)
\put(0,0){
\includegraphics[clip]{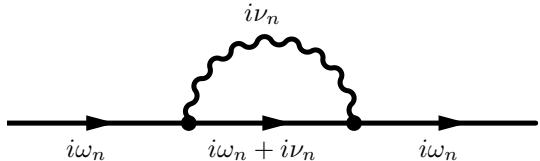}
}
\fmfL(33.37534,44,t){$i\omega _n$}%
\fmfL(100.00014,44,t){$i\omega _n+i\nu _n$}%
\fmfL(100,87.24988,b){$i\nu _n$}%
\fmfL(166.6247,44,t){$i\omega _n$}%
\end{picture}

\caption{Feynman diagram of the electron-phonon self-energy $\Sigma^{\rm ph}(z)$
in $\lph^2$, including the two attached incoming and outgoing electron-propagator lines. 
The full line represents the local Green function $G_d(z)$, the wiggled line
the phonon-propagator.
}
\label{fig:self-electron-phonon-diagram}
\end{figure}

In the weak coupling regime, $g^2\ll 1$, the spectral function is only
slightly altered by the electron-phonon coupling. Starting from the
free Green function of the RLM, the second-order contribution to the
self-energy is  given by the Feynman diagram shownin Fig.\
\ref{fig:self-electron-phonon-diagram}. Evaluating this leading order
text-book diagram\cite{Mahan81}  yields the  analytical expression 
\begin{eqnarray}
\label{eqn:self-energy-diagram-e-ph}
\Im m \Sigma^{\rm ph}(\w-i\delta) &=&
\pi\lph^2[g(\w_0)   +f(\w+\w_0) ] \rho_d(\w+\w_0) \non
&& +\pi\lph^2[g(\w_0) +f(\w_0-\w)]\rho_d(\w-\w_0) \non
\end{eqnarray}
where $f(\w)$ ($g(\w)$) is the Fermi (Bose) function. The self-energy
essentially vanished for $|\w|<\w_0$ and $T\to 0$ and acquires a
significant value only for $|\w|> \w_0$. The sharpness of the increase
is governed by the Fermi function:  an energy quantum of at least one
bosonic excitation energy $\w_0$  is needed to open up the additional
decay channel.  

In a conserving approximation, one would replace the phonon propagator
by its fully dressed version which includes the shift of the phonon
frequencies as well as a finite life-time broadening due to the local
particle-hole excitations. This effect of the charge susceptibility
has been neglected in Eq.\ (\ref{eqn:self-energy-diagram-e-ph}). If we
include this effect we would need to convolute  $\Im m \Sigma^{\rm
  ph}(\w-i\delta)$ with the  true phonon spectral function which has
been replaced by $\delta$-functions in the derivation of Eq.\
(\ref{eqn:self-energy-diagram-e-ph}). Therefore, we expect  (i) a
broadening of the steep increase above $\w_0$ with increasing $\lph$
as well as additional features at multiples of the phonon frequency,
since $ \Sigma^{\rm ph}(\w-i\delta) $ will also enter
self-consistently the  charge susceptibility. 

\begin{figure}[t]
\begin{center}
\includegraphics[width=85mm]{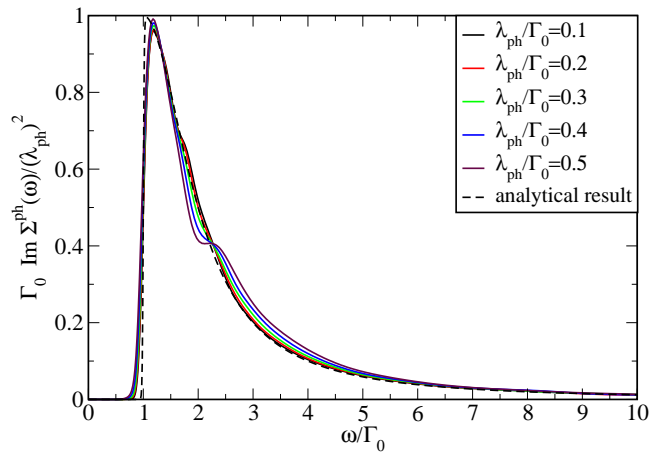}

\caption{(Color online.) Imaginary part of self-energy contribution $\Sigma^{\rm ph}(\w-i\delta)$ for $E_d=0$, $\w_0=\Gamma_0$ and various values of $\lph$ in the
weak coupling regime $\lph/\Gamma_0<1$ and positive frequencies. The dashed line depicts
the analytic self-energy given by
 Eq.\ (\ref{eqn:self-energy-diagram-e-ph}) and  evaluated for the same temperature 
$T=8.5\times 10^{-3}\Gamma_0$
as the NRG results.
NRG parameter: $N_s=1500$, $\Lambda=1.6$, $N_b=40$, $N_z=8$, $b=0.1$.
}
\label{fig:sigma-weak-coupling-ed0-w0-1}
\end{center}
\end{figure}

In Fig.\ \ref{fig:sigma-weak-coupling-ed0-w0-1}, the self-energy
contribution $\Im m \Sigma^{\rm ph}(\w-i\delta)$ close to the real
axis is depicted for few values  $(\lph/\Gamma_0)^2\ll 1$. Dividing
out the leading order prefactor, $\lph^2$,  the magnitude of
self-energy becomes nearly independent of $\lph$. The analytical
expression of Eq.\ (\ref{eqn:self-energy-diagram-e-ph}) added as
dashed line shows a remarkably good agreement with the NRG self-energy
obtained via  Eq.\ (\ref{eq:nrg-self-energy}). Due to the  broadening,
the NRG self-energy cannot generate the same steep increase of the
self-energy at $\w\approx \w_0$.  Note, that a shoulder is slowly
developing at $\w\approx 2\w_0$ with increasing $\lph$, a precursor of
additional phonon-peaks in the spectra. This is due to higher order
processes not included in the second-order perturbation expansion, but
present in a proper conserving approximation  for $\Sigma^{\rm ph}(z)$.

\subsection{Particle-hole asymmetry}

Recently, H\"utzen et al.\ \cite{HutzenEgger2012} investigated the I-V
characteristics of the spin-less Anderson-Holstein model  using an
iterative real-time path integral approach.\cite{Egger08} Within this
approach   the current is directly calculated using a generating
functional, and  no spectral functions are required. Signatures of a
Franck-Condon blockade have been  reported using the local  molecular
Hamiltonian $H_m$ 
\begin{eqnarray*}
H_{m}&=&  \w_0  b^\dagger  b + \lambda_{ph}  
( b^\dagger + b)  d^\dagger d  +\e d^\dagger d
\end{eqnarray*}
defined in Eq.\ (\ref{eqn:Hm-egger})  for  the parameters $\e=0$ and
$\w_0=2\Gamma_0$: the suppression of the current\cite{HutzenEgger2012}
has been found when increasing  the electron-phonon coupling
$\lph$. Since $H_{imp}$  is identical  to  $H_m$  by setting $n_0=0$
and $E_d = \e- E_p$,  the suppression of the current at zero bias can
be partially understood  in terms of the polaronic shift  of the
single-particle energy $E_d= \e-g^2\w_0$.  Increasing $\lph$ decreases
$E_d$  for constant $\e$ and, therefore, increases the particle-hole
asymmetry. As a consequence, the stronger the electron-phonon
coupling, the more energy is required to depopulate the level.

\begin{figure}[t]
\begin{center}
\vspace*{5mm}
\includegraphics[width=85mm]{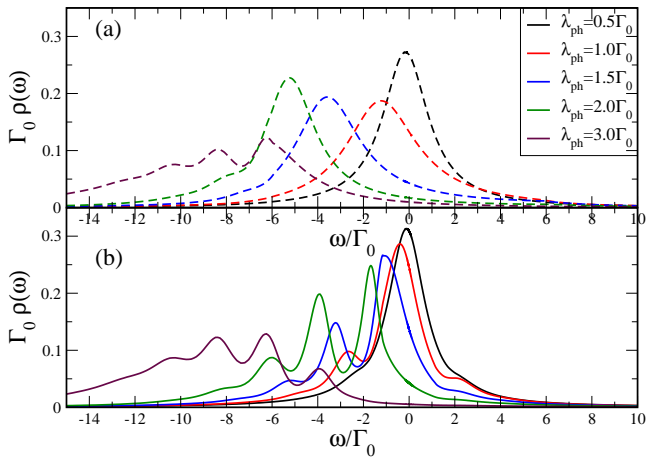}

\caption{
(Color online.) 
Equilibrium spectral function for the molecular impurity Hamiltonian $H_m$, Eq.\ (\ref{eqn:Hm-egger})
from weak to strong coupling in the anti-adiabatic regime $\w_0=2\Gamma_0$.
The solid lines in (b) show spectra at  $T=0.2\Gamma_0$, the dashed lines in (a)
the spectra for  $T=1.2\Gamma_0$. With $\e=0$ and  $n_0=0$, $E_d$ is given by $E_d=-g^2w_0$ 
in $H_{imp}$.
NRG parameter: as in Fig.\ \ref{fig:sigma-weak-coupling-ed0-w0-1}.
}
\label{fig:franck-condon-spectra}
\end{center}
\end{figure}

The spectral functions   are shown  for
two different temperatures and $\e=0$ and $\w_0=2\Gamma_0$ in Fig.~\ref{fig:franck-condon-spectra}. 
With increasing $\lph$, the main resonance peak is shifted to lower energies by $E_p$. 
Additional phonon-side peaks
occur  asymmetrically  around the main resonance at $E_d$. The decrease
of the spectra at $\w=0$ is clearly visible translating immediately  to the 
reported\cite{HutzenEgger2012} suppression of the zero bias conductance. 
By comparing Fig.~\ref{fig:franck-condon-spectra}(a) and (b), we also note
an increasing shift of the spectra toward the chemical potential and
a visible narrowing of the main resonance when lowering the temperature for fixed $\lph$.

%
%
%
%
%
%
%

\subsection{Benchmark for non-equilibrium spectral functions}
\label{sec:benchmark}

The scattering states NRG approach to quantum
transport\cite{AndersSSnrg2008, SchmittAnders2010,SchmittAnders2011}
relies on the analytically known density operator of a non-interacting
model at finite bias.\cite{Hershfield1993} It is evolved to the
density-matrix of the fully interacting problem using the
TD-NRG.\cite{AndersSchiller2005,AndersSchiller2006} This
non-equilibrium density operator for  the limit $t\to\infty$  is used
to calculate the  finite-bias retarded spectral function entering  the
equation  (\ref{eq:ss-current}) for the current-voltage characteristic
of the junction.\cite{MeirWingreen1992}

The restriction to the single-lead version of the  Hamiltonian defined
in Eq.\ (\ref{eqn:two-lead-RLM})  in the absence of a bias can be used
to benchmark the quality  of the non-equilibrium spectra function
algorithm and its limits.  Starting from a decoupled electron-phonon
system, $\lph=0$,  and letting the system evolve using the full
Anderson-Holstein model  $H=H_0+H_{ph}$,  we must recover the
equilibrium dynamics of $H$ for $t\to\infty$.  Therefore, we compare
the results  for the spectral function obtained from the standard
equilibrium NRG algorithm\cite{PetersPruschkeAnders2006} with the
spectra obtained from a  extension of the TD-NRG to spectral
functions.\cite{AndersNeqGf2008} 
\begin{figure}[tb]
\begin{center}
\vspace*{5mm}
\includegraphics[width=85mm]{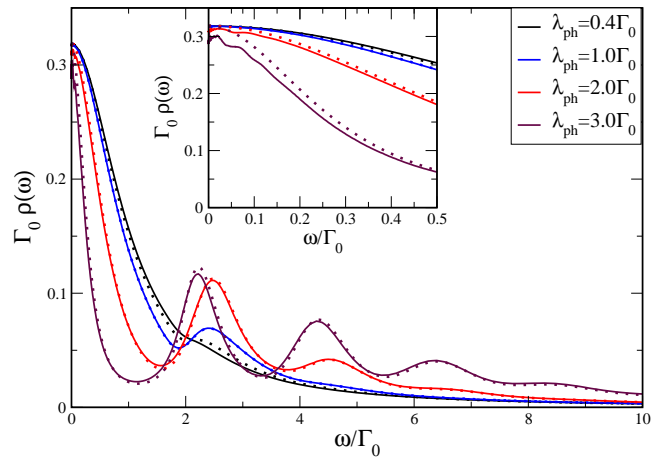}

\caption{
(Color online.) 
Comparison between the equilibrium and non-equilibrium spectral
function for a single lead and the phonon energy $\w_0=2\Gamma_0$. The
dotted lines show the equilibrium spectral function,  the solid lines
with the same color the spectra obtained for times $t\to\infty$ after
a quench by switching on the electron-phonon coupling at time $t=0$. 
NRG parameter: $N_s=1500$, $\Lambda=1.6$, $N_b=40$, $N_z=8$, $b=0.1$, $T=8.5\times 10^{-3}\Gamma_0$.
}
\label{fig:eq-vs-neq-spectral-w0-2}
\end{center}
\end{figure}

The results for $\w_0=2\Gamma_0$, $E_d=0$ and
various electron-phonon coupling constants $\lph$ are depicted in 
Fig.\ \ref{fig:eq-vs-neq-spectral-w0-2}. The dotted line shows the equilibrium 
spectral function, the solid line the spectra obtained
from a quench by switching on the electron-phonon coupling at $t=0$. The agreement remains
very good even at high energies up to moderate values of $\lph$.

The larger the phonon frequency, the better the agreement between the
direct calculation of equilibrium spectra and the spectra obtained by
the non-equilibrium approach. For small $\w_0\le\Gamma_0$, however,
the non-equilibrium approach has its limitations. The number of
bosonic Fock states which are required even in equilibrium grows
significantly. For our equilibrium calculation, we needed already
$N_b=400$ states, in a recent study,\cite{EidelsteinSchiller2012} the
rather very generous choice of NRG parameter, $N_b=3000$ and
$N_s=8000$,  have been used in a single lead model but no z-averaging
was required for the thermodynamical properties investigated in that
paper. For the required z-averaging over 64 discretizations in an
TD-NRG calculation, this would be computationally much to expensive
and is, therefore, beyond the reach of our approach.

\section{Quantum transport: Results}
\label{sec:quantum-transport}

\subsection{Scattering-states NRG}

This section is devoted to the results of the non-linear I-V
characteristics  calculated using the SNRG. The differential
conductance $G(V)=dI/dV$ has been obtained  numerically from the I(V)
curves. 

Section \ref{sec:q-transport-ph-sym}, is devoted to the particle-hole
symmetric regime of the model, i.\ e.\ $E_d=0$. We investigate the
evolution of the I-V characteristics from a junction with symmetric
couplings ($R=1$)  to the tunneling regime ($R\to\infty$.)  In Sec.\
\ref{sec:ph-asymmetric-regime} we focus on the PH-asymmetric model and
investigate the Franck-Condon blockade for $\e=E_d- E_p=0$ and a
finite electron-phonon coupling.

\subsubsection{Particle-hole symmetric model}
\label{sec:q-transport-ph-sym}

\begin{figure}[tb]
\begin{center}
\vspace*{5mm}

\includegraphics[width=85mm]{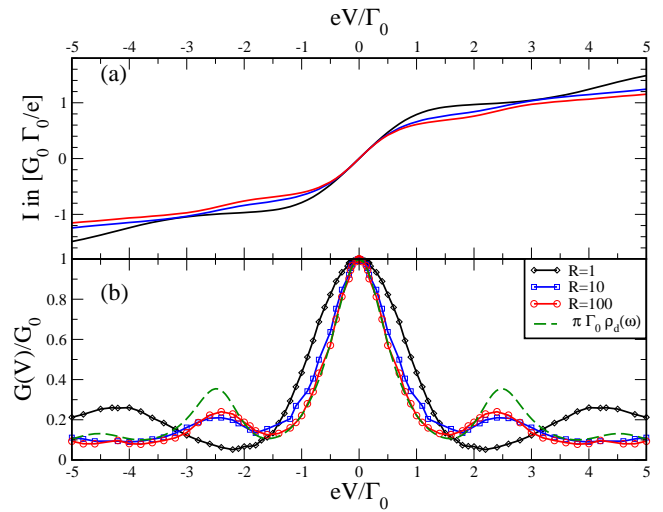}

\caption{
(Color online.) 
(a) I(V)  for the symmetric model, i.\ e.\ $E_d=0$, and three
different ratios $R=\Gamma_L/\Gamma_R =1,10,100$. The vibrational
parameters are $\w_0=\lph=2\Gamma_0$.
(b) $dI/dV$  in units of $G_0$ as function of $eV$ obtained by
numerically differentiating the curves in (a). Additionally, the
equilibrium transmission function $T(\w)=\pi\Gamma_0\rho_d(\w)$
obtained from the spectral function $\rho_d(\w)$ of Fig.\
\ref{fig:eq-vs-neq-spectral-w0-2} has been added for comparison. 
NRG parameter: $N_s=2000$, $\Lambda=2$, $N_b=35$, $N_z=64$, 
$b=0.15$, $T=6.5\times 10^{-3}\Gamma_0$.
}
\label{fig:dI-dV-w0-2-lph=2}
\end{center}
\end{figure}

The electrical current calculated via Eq.\ (\ref{eq:ss-current})  is
shown as a function of the applied bias  in
Fig.~\ref{fig:dI-dV-w0-2-lph=2}(a)  for three different ratios of
$R=\Gamma_L/\Gamma_R$, a phonon frequency $\w_0=2\Gamma_0$  and a
fixed electron-phonon coupling $g=1$. For better comparison, the
prefactor $G_0$ defined in Eq.\  (\ref{eqn:G0-prefactor}), has been
divided out. $G_0$  only contains the trivial reduction of the current
upon increasing  of $R$ starting from its maximum of $e^2/h$ for
$R=1$.  For a PH-symmetric Hamiltonian, the current remains
PH-symmetric  even for an asymmetric junction as can be seen
analytically from Eq.\ (\ref{eq:ss-current}). 

The differential conductance $G(V)=dI/dV$ depicted in
Fig.~\ref{fig:dI-dV-w0-2-lph=2}(b) has been obtained by numerically
differentiating the data of Fig.~\ref{fig:dI-dV-w0-2-lph=2}(a). For a
better comparison, we added the corresponding equilibrium transmission
function  obtained from the spectral function shown in the Fig.\
\ref{fig:eq-vs-neq-spectral-w0-2}.

Particle-hole symmetry and a Fermi-liquid equilibrium ground state
yields a  pinned peak at zero bias which approaches $G_0$ for $T\to 0$
which can be understand quite easily by applying the Friedel sum
rule.\cite{Langreth1966,Anders1991} The narrowing of the central
resonance due to the renormalization of $\Gamma_0\to\Gamma_{\rm eff}$,
exemplified in Fig.\ \ref{fig:eq-vs-neq-spectral-w0-2} prevails also
in the differential conductance. 

Qualitatively, $G(V)$ traces the transmission function
$T(\w)=\pi\Gamma_0\rho(\w)$ but quantitatively significant differences
are observed. The additional transmission maxima are clearly visible
at finite voltage which are related to phonon-assisted tunneling
increasing significantly the current above a threshold. This threshold
position depend  on the asymmetry $R$ of the  junction.

We expect that  $G_0^{-1}dI/dV$  approaches the equilibrium
transmission function in the tunneling regime  $R\to \infty$ and $T\to
0$. This is clearly the case for $R=100$ and low voltages. We note,
however, that the peak position of first maxima at $eV\approx \pm
\w_0$  is well reproduced but the peak height  is smaller than
expected from the equilibrium transmission function
$T(\w)=\pi\Gamma_0\rho_d(\w)$. This might be caused by remaining
non-equilibrium effects still  relevant for $R=100$. However,  we
believe that this is cause by the limitations of the numerical
accuracy of the SNRG at large bias caused by the discretisation and
the broadening of the spectral function in combination with the TD-NRG
time evolution. 

By decreasing the asymmetry from $R=100$ to $R=10$ small differences
are visible which can be traced back mainly to the changes chemical
potentials $\mu_L$ and $\mu_R$ for the same  voltage drop. For a
symmetric junction $R=1$, the strongest non-equilibrium effects are
observed. $G(V)$ does not follow the expectation   $G(V)\propto
[T(eV/2) +T(-eV/2)]$ where  the equilibrium transmission function
$T(\w)$ naively has replaced full bias dependent spectra
$\rho_d(\w,V)$ in  Eq.\ (\ref{eq:ss-current}). The width of the
central peak is smaller that expected for $G(V)\propto [T(eV/2)]$ and
the peaks from the phonon assisted tunneling occur closer to $\pm
2\w_0$ as suggested by the equilibrium transmission function. This is
due to a significant change of the non-equilibrium spectral function
with increasing bias voltage. Spectral weight is redistributed to
higher frequency due to backscattering processes for which additional
phase space becomes available at finite voltage. 

The full width at half maximum (FWHM) of the zero bias $dI/dV$ peak is
not simply given by $2\Gamma_{\rm eff}$  as suggested by a simple
generalization of equilibrium spectral function. Non-equilibrium
effects cause a bias dependency of the spectral function:  the width
of $dI/dV$ is significantly smaller than predicted from such a  naive
$[T(eV/2) +T(-eV/2)]$ fit to $G(V)$ for $R=1$.

\subsubsection{Particle-hole asymmetric model}
\label{sec:ph-asymmetric-regime}

\begin{figure}[tb]
\begin{center}

\includegraphics[width=85mm]{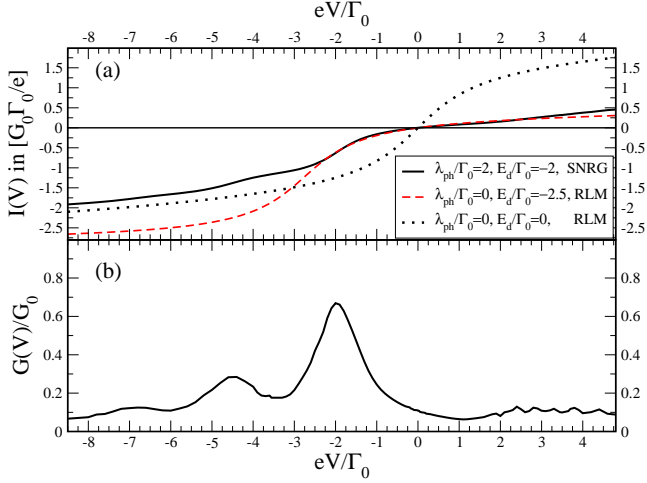}

\caption{
(Color online.) 
(a) I(V)  for the particle-hole asymmetric model, $E_d/\Gamma_0=-2
(\e=0)$ and an asymmetric junction $R=\Gamma_L/\Gamma_R =10$. The
vibrational parameter are $\w_0=2\Gamma_0$, and $\lph=2\Gamma_0$.  The
solid line shows the  SNRG current, The black dotted line indicates
the $\lph=\e=0$ current, the blue dashed line current for a shifted
level $E_s/\Gamma_0=-2.5$ in an non-interacting junction (RLM)at
$T=0$. 
(b) $dI/dV$  in units of $G_0$ as function of $eV$ obtained by
numerically differentiating the SNRG curve in (a).
NRG parameter: as in Fig.\ \ref{fig:dI-dV-w0-2-lph=2}.
}
\label{fig:dI-dV-eps=0-R=10-w0-2-lph=2}
\end{center}
\end{figure}

\begin{figure}[tb]
\begin{center}

\includegraphics[width=85mm]{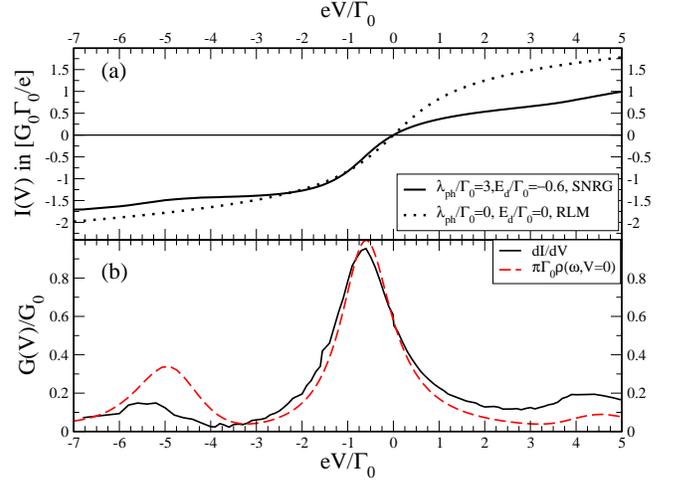}

\caption{
(Color online.) 
(a) I(V)  for  $E_d/\Gamma_0=-0.6 (\e/\Gamma_0=1.65)$ and
$R=\Gamma_L/\Gamma_R =10$. The vibrational parameter are
$\w_0=4\Gamma_0$, and $\lph=3\Gamma_0$.  The solid line shows the
SNRG current, The black dotted line indicates the $\lph=\e=0$ current
(RLM) at $T=0$. 
(b) $dI/dV$  in units of $G_0$ as function of $eV$ obtained by
numerically differentiating the SNRG curve in (a). 
NRG parameter: as in Fig.\ \ref{fig:dI-dV-w0-2-lph=2}.
}
\label{fig:dI-dV-Ed=-0.6=R=10-w0-4-lph=3}
\end{center}
\end{figure}

For a particle-hole asymmetric regime of the model and asymmetric lead
couplings, the current as function of the applied bias voltage and
corresponding  the differential conductance $G(V)$ is depicted in
Fig.\ \ref{fig:dI-dV-eps=0-R=10-w0-2-lph=2}.  We have set $\e=0$ in
$H_m$ in resonance with both chemical potentials at zero bias, the
phonon frequency to $\w_0/\Gamma_0=2$ and selected a moderate
electron-phonon coupling $g=\lph/\w_0=1$. As discussed before, the
bare single-particle  level $\e$ is shifted by the polaron energy to
$E_d=-E_p=-2\Gamma_0$.  The  corresponding equilibrium spectral
function has been plotted in Fig.\ \ref{fig:franck-condon-spectra}
which clearly documents  the significant asymmetry.

The strong suppression of the current with increasing $\lph$ and fixed $\e$,
as seen by comparing the $\lph=0$ and $\lph/\Gamma_0=2$ curves has
been interpreted as Franck-Condon blockade
physics.\cite{HutzenEgger2012}   Clearly, the leading order effect
stems from a polaronic shift $E_p$ of the single-particle level. The
SNRG curve  tracks the current through a  shifted level at
$E_d/\Gamma_0=-2.5$ and $\lph=0$ rather well for positive voltages as
indicated by the read dashed line.

We have already seen in the equilibrium spectral function -- green
curve in Fig.\ \ref{fig:franck-condon-spectra} -- that the
electron-phonon interaction causes a redistribution of spectral weight
below the chemical potential. Each phonon peak contains only of a
fraction of the spectral weight, and  spectral weight has been shifted
even below $-10\Gamma_0$ which contribute only at much larger negative
voltages. In the asymmetric junction ($R=10$), the current $|I|$
increases significantly for $eV<-E_p$.  However, the magnitude of
current  clearly remains lower than the reference current (dashed
line) for a non-interacting shifted level.  This is consistent with
the the qualitativ features of the equilibrium spectra and leads to
the Franck-Condon current suppression of the current at small and
intermediate bias voltage. A second phonon induced maximum in $G(V)$
is found below $V < -E_p-\w_0$ and corresponds to the second peak in
the spectral function.  A very shallow maximum at $eV/\Gamma_0\approx
-7$ could be related to the third phonon peak in the spectral
function. We believe, that the small oscillations in die $dI/dV$ curve
for positive bias voltage has no physical significance and is caused
by the numerical differentiation of the numerical I-V data.

Maintaining the same coupling asymmetry $R=10$ but increasing the
phonon energy to $\w_0/\Gamma_0=4$, the current vs voltage is depicted
in Fig.\ \ref{fig:dI-dV-Ed=-0.6=R=10-w0-4-lph=3}(a) and the
corresponding $dI/dV$ curve in Fig.\
\ref{fig:dI-dV-Ed=-0.6=R=10-w0-4-lph=3}(b) for a moderate electron
phonon coupling $g=0.75$. We also added the equilibrium spectral
function to the (b) as a comparison. Since $R=10$, $d\rho(\w,V)/dV$
remains significant, and  $dI/dV$ does not trace the equilibrium
spectra $\rho(\w,V=0)$. There remains a significant renormalization of
the spectral function  for a moderate asymmetry which only must vanish
in the limit $R\to \infty$ or $R\to 0$.

\subsection{Tunneling limit for the crossover regime}

In the adiabatic regime the physics is dominated by the
charge-fluctuation scale $\Gamma_0$ being the largest local energy
scale in the problem. The weak electron-phonon coupling causes only
small renormalization of the electronic and phononic degrees of
freedom. In this regime, Keldysh approaches  have been successfully
applied to the problem: the  phonon frequency is reduced by
particle-hole excitations in leading order which is well captured
already in conserving one-loop
diagrams\cite{Caroli71,Caroli72,GalperinRatnerNitzan2007} for the
electron and the phonon propagator. 

Recently, it was shown\cite{EidelsteinSchiller2012} that the
anti-adiabatic regime extends from $\w_0>\Gamma_0$ to $\w_0 \approx
O(\Gamma)$, as long as the polaronic shift $E_p$ exceeds the
charge-fluctuation scale. The  crossover regime from  this extended
anti-adiabatic regime to the adiabatic regime  where the polaron
contains less and less phononic excitations has been
defined\cite{EidelsteinSchiller2012} by the parameter hierarchy
$E_p>\Gamma_0> \w_0$ and $\Gamma_{\rm eff}\approx \w_0$.

Although this regime is accessible to  the equilibrium NRG, a very
large number  of  phononic states are needed for accurately tracking
the equilibrium flow.\cite{EidelsteinSchiller2012} The SNRG relies on
the  switching on the electron-phonon coupling and let the system
evolve from a  $\lph=0$ to a finite $\lph$. Apparently,  the TD-NRG is
not able to reproduce reliably  the equilibrium Green functions for
larger electron-phonon couplings in the crossover regime. This might
also be  related to the logarithmic discretisation of the scattering
states  in combination with  the change of the ground
states\cite{EidelsteinGuettgeSchillerAnders2012,GuettgeAndersSchiller2012} 

Therefore, we have focused on the tunneling regime ($R\to\infty$)
where  the retarded Green function becomes bias independent and
$\rho_d(\w,V)$ can be replaced in Eq.\ (\ref{eq:ss-current}) by
$\rho_d(\w,V=0)$.  We have included $N_b=300$ phonon Fock states and
kept $N_s=2500$ NRG after each iteration to calculate the equilibrium
spectra using only a  single-lead.

\begin{figure}[bt]
 \includegraphics[angle=270,width=85mm]{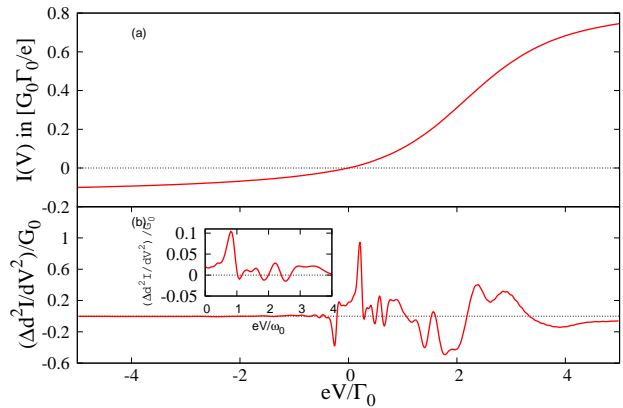}
\caption{(a) I(V) for the particle-hole asymmetric model,
$E_d/\Gamma_0=1.2$ in the crossover regime. The phonon energy
$\omega_0 = 0.26 \Gamma_0$, and $\lambda_{\mathrm{ph}} = 0.6
\Gamma_0$. (b) $\Delta\mathrm{d}^2 I/\mathrm{d} V^2$ in units of $G_0$
as a function of eV obtained  by taking  analytically the second
derivative of Eq.\ (\ref{eq:ss-current}) in the limit $R\to\infty$. 
NRG parameter: $ N_\mathrm{s} = 2500, \Lambda = 1.8, N_\mathrm{b} =
300, N_\mathrm{z} = 8, b = 0.1, T = 8 \times 10^{-4} \Gamma_0$. 
} 
\label{fig:13}
\end{figure}

Since the charge-fluctuation scale $\Gamma_0$ is the dominating energy
scale, $G(V)$ is unsuitable to reveal  subtle inelastic processes
induced by the electron-phonon coupling in the cross-over regime. The
information about the inelastic processes,  however, is encoded in the
second derivative\cite{Caroli71,Caroli72,GalperinRatnerNitzan2007} of
the $I(V)$ curve,  $\mathrm{d}^2 I/\mathrm{d} V^2$.  

For the tunneling  $I(V)$ is depicted in  Fig.\
\ref{fig:13}(a). Although, $\w_0<\Gamma_0$, the  polaron shift is
given by $E_p/\Gamma_0= 1.38$ and exceeds $\Gamma_0$,  while the
renormalized $\Gamma_{\rm eff}$ remains above $\w_0$. Therefore, the
choice of parameters lies in the crossover regime between the extended
anti-adiabatic and the adiabatic regime. 

Since $\mathrm{d}^2 I/\mathrm{d} V^2$  is up to some smaller voltage
dependence   proportional to the derivative  of the spectral function,
it is sensitive to the derivative to the phonon self-energy. Although,
this absolute magnitude is small, its derivative can  be huge if it
contains a threshold set by a Fermi-function shifted by one or multiples
of the phonon frequencies. In  second order Keldysh approximation,
these steep features dominate $\mathrm{d}^2 I/\mathrm{d} V^2$  on
voltages $eV/\w_0\approx O(1)$ as seen in Fig.\ 11 of  Ref.\
\onlinecite{GalperinRatnerNitzan2007}. The sign change of the
$\mathrm{d}^2 I/\mathrm{d} V^2$  for small voltages can only occur for
very pronounce changes in the self-energies.

In the NRG, the features are much less dominant: the method has its
limitation due to the discretization and the artificial broadening
which puts limit to the gradient of any self-energy change as already
discussed above.  In order to extract some information about inelastic
scattering processes, we have calculated $\mathrm{d}^2 I_0/\mathrm{d}
V^2$ by using a Lorentz fit to the spectral function centered around
$\bar E_d$ which is the mean-field (MF) reference value. Now we
calculate the full $\mathrm{d}^2 I/\mathrm{d} V^2$ and define $\Delta
\mathrm{d}^2 I/\mathrm{d} V^2 =\mathrm{d}^2 I/\mathrm{d} V^2
-\mathrm{d}^2 I_0/\mathrm{d} V^2$ as difference between the full
second derivative and the MF result. This is plotted in Fig.\
\ref{fig:13}(b). Now we notice the sharp features around $eV=\pm \w_0$
and at multiples of the phonon frequencies. This is clearly visible in
the inset of  Fig.\ \ref{fig:13}(b) where the bias is measured in
units of $\w_0$. While for $\lph\to 0$, the self-energy is dominated
by threshold of the inelastic processes at $|\w|=\w_0$, for increasing
$\lph$ multiple phonon absorption and  emission processes start to
contribute to the self-energy. Similar to Fig.\
\ref{fig:sigma-weak-coupling-ed0-w0-1} where two phonon processes are
clearly visible, this multi-phonon processes at moderate coupling
yield the additional features in $\Delta \mathrm{d}^2 I/\mathrm{d}
V^2$. The one and two-phonon processes have been also seen in a much
more pronounce manner  in the
literature\cite{GalperinRatnerNitzan2007} using Keldysh perturbation
theory.

\section{Discussion and outlook}
\label{sec:discussion-outlook}

We have extended the scattering states NRG to the  charge-transport
through a molecular junction. To set the stage for the non-linear
transport results, we have provided a detailed analysis of the
low-temperature equilibrium physics of the spin-less Anderson-Holstein
model from an NRG perspective. We have shown that the model can been mapped
onto an interacting resonant level model in the extended
anti-adiabatic regime and extracted the  renormalized charge-transfer
scale $\Gamma_{\rm eff}$ and the effective  Coulomb interaction
$U_{\rm eff}$ in a regime complementary to the recent study of
Eidelstein et al..\cite{EidelsteinSchiller2012} 

We have calculated the equilibrium spectral functions and used those to
benchmark our non-equilibrium algorithm.\cite{AndersNeqGf2008} We have
demonstrated that for the weak coupling limit,  $(\lph/\w_0)^2\ll 1$,
the NRG tracks perfectly the self-energy obtained from the
second-order Feynman diagram. For large electron-phonon couplings the
typical phonon replica peaks of the exact  atomic
solution\cite{LangFirsov1962,Mahan81}  are found. Using extensive
z-averging  the spectra become independent of the NRG broadening even
at higher frequencies.  
 
Exemplified in Fig.~\ref{fig:dI-dV-w0-2-lph=2}, the reduction of the
charge-transfer scale $\Gamma_{\rm eff}$  due to the electron-phonon
coupling in the equilibrium properties conveys into a narrowing of the
differential conductance.  In contrary to a recent QMC
study\cite{HanMolecule2010} for the spin-full model with Coulomb
interaction, our results clearly show the phonon side peaks  expected
from the equilibrium spectra. The location, however, depends on the
two chemical potentials as well as the coupling  asymmetry of the
junction. 

Gating the junction away from the particle-hole symmetric point
reveals the Franck-Condon blockade physics with increasing
electron-phonon coupling.  The leading order effect is understood in
terms of a polaronic level shift, and  the I(V) curve tracks a
shifted resonant level model for positive bias voltages. For negative
bias, however, the current is suppressed due to a redistribution of
spectral weight to higher frequencies.

\begin{acknowledgments}
We are grateful to Jong E.~Han, Holger Fehske,
Avraham Schiller for helpful discussions.
We are particularly grateful to Avraham Schiller, who sent us
a preprint of Ref.\ \onlinecite{EidelsteinSchiller2012}.
This work was supported by the German-Israeli
Foundation through grant no.\ 1035-36.14,
by the Deutsche Forschungsgemeinschaft under AN 275/6-2,
and supercomputer support was provided by the NIC, FZ J\"ulich under project
No. HHB00.

\end{acknowledgments}


%

\end{document}